\documentclass[twocolumn,10pt,floatfix,superscriptaddress,aps,prl]{revtex4-2}
\pdfoutput=1
\usepackage[utf8]{inputenc}
\usepackage[T1]{fontenc}
\usepackage[normalem]{ulem}
\usepackage[pdftex]{color,graphicx}
\usepackage{amsmath, amssymb, mathtools}
\usepackage{bm}
\usepackage{dsfont}
\usepackage{color}
\usepackage{leftindex}
\usepackage{float}
\usepackage[final]{pdfpages}
\usepackage{lipsum}
\usepackage[colorlinks=true,linkcolor=red,anchorcolor=red,citecolor=blue,urlcolor=blue]{hyperref}

\newcommand{\DC}{\textrm{DC}}
\makeatletter

\AtBeginDocument{\let\LS@rot\@undefined}
\makeatother

\begin{document}
\title{Physical constraints on effective non-Hermitian systems}

\author{Aaron Kleger}
\email{aaron.r.kleger.gr@dartmouth.edu} 

\author{Rufus Boyack}
\email{rufus.boyack@dartmouth.edu}
\affiliation{Department of Physics and Astronomy, Dartmouth College, Hanover, New Hampshire 03755, USA}

\date{\today}

\begin{abstract}
Interacting and open quantum systems can be formulated in terms of an effective non-Hermitian Hamiltonian (NHH), however, there are important constraints that must be satisfied by the effective action and the associated Green's functions. One common approach to many-body non-Hermitian (NH) systems is to incorporate the anti-Hermitian part of the Hamiltonian directly in the Matsubara Green's function. Here, we show that such an approach is incompatible with the standard framework for systems with interactions. Furthermore, we furnish a consistent physical description for such systems by determining their distinction from conventional interacting physics, and find that they are described by pseudo-Hermitian quantum mechanics. In addition, we characterize the zero-temperature distribution functions within several frameworks for NH systems. As an application of our results, we consider the electromagnetic response of a NH quasiparticle Hamiltonian based on the (1+1)-dimensional NH Dirac model subject to various physical descriptions.
\end{abstract}
\maketitle

\section{Introduction} Non-Hermitian (NH) systems~\cite{Bender1998, Bender1999, Bender2007} have attracted significant interest in recent years due to the existence of an abundance of exotic phenomena, including skin effects~\cite{Okuma2020}, enhanced sensitivity~\cite{Wiersig2014, Hashemi2022}, unique topological phases~\cite{Borgnia2020, Shen2018, Papaj2019, Nagai2020, Rivero2023} and quantum-critical phenomena~\cite{Nakagawa2018}. While standard approaches to quantum mechanics require the Hamiltonian to be Hermitian, the interactions between a particle and its environment can give rise to a quasiparticle description in terms of an effective non-Hermitian Hamiltonian (NHH)~\cite{Kozii2024, GomezLeon2022, Thompson2023, Michishita2020, Meden2023}. The self-energy of the single-particle Green's function encapsulates the effects of these interactions, without the requirement of being a Hermitian operator~\cite{Kamenev_2023}. Although the requirement of Hermiticity is relaxed for the quasiparticle Hamiltonian, there still exist constraints on the form of the self-energy to remain consistent with the standard framework for interacting systems~\cite{AGD, Kamenev_2023}. 

There exist numerous mathematical frameworks for computing the physical properties of a NH system, such as describing an open quantum system (OQS)~\cite{Manzano2020, BreuerBook, Lindblad1976, Gorini1976} in terms of its dynamics conditioned on the absence of quantum jumps~\cite{Sticlet2022, Pan2020, Meden2023, Ashida2018, Herviou2019, Roy2023, Kou2022, Cao2023b, Cao2023, Soares2025, Michishita2020, Clerk2022}, the use of biorthogonal quantum mechanics (BQM)~\cite{Brody2013}, parity-time ($\mathcal{PT}$)-symmetric~\cite{Bender2024} or pseudo-Hermitian quantum mechanics (PHQM)~\cite{Mostafazadeh2010}, and Keldysh field theory~\cite{Sieberer2016, Kamenev_2023, Thompson2023, Talkington2024}. Formally, the dynamics of any of these approaches can be encapsulated in terms of their Green's functions. For example, the retarded Green's function is
\begin{equation}
G_{R}(\omega) = (\omega - H)^{-1},
\end{equation}
where $H \equiv H_0+ \Sigma_{R}$ is an effective NHH; note that the scalar quantities are implicitly accompanied by the identity operator $\mathds{1}$. Taking $H_0$ as the bare Hermitian Hamiltonian allows the anti-Hermitian part of the effective Hamiltonian to be interpreted as part of the self-energy. Regardless of the approach, physical constraints, such as causality and gauge invariance of observables, still apply. Note that we use Natural units $\hbar=k_{B}=1$. 

A popular approach to NH physics is described by PHQM~\cite{Mostafazadeh2010}. For a NHH that satisfies $H^{\dagger} = \eta H \eta^{-1}$ with a Hermitian operator $\eta$, unitary time evolution is recovered under a modified inner product. A NH system can be directly related to an isospectral Hermitian system by redefining the inner product in terms of $\eta$. Although PHQM does not provide a direct motivation for NHHs, it offers an alternative framework and treatment for them.

Meden \textit{et al}.'s recent review~\cite{Meden2023} emphasized that with the introduction of concepts from $\mathcal{PT}$-symmetric QM, BQM, and PHQM, there has emerged a lack of consensus on central issues such as how expectation values are to be computed and what form equilibrium density matrices should take. The differences in the dynamics of these approaches have been highlighted in Refs.~\cite{Meden2023, Herviou2019, Sim2024}. The association of the use of right-right or right-left eigenstates to define the density matrix and expectation values with the physical setup has been suggested~\cite{Groenendijk2021}. However, the use of biorthogonal eigenstates is concomitant with any diagonal spectral representation of finite-dimensional NH systems and does not necessarily imply the use of BQM or PHQM. 

In order to study many-body interacting systems, finite-temperature techniques with a consistent physical picture must be established. For many works, the source of the non-Hermiticity is phenomenological, instead of resulting from a direct calculation of the self-energy. Nevertheless, physical constraints on the non-Hermiticity in the system must still be imposed. For example, it is conventionally understood (that is, without modifying the inner product) that the causal Green's functions are related under Hermitian conjugation: $G_A=(G_R)^\dagger$. Many works define finite-temperature Green's functions which do not satisfy this constraint, which leaves open the question as to how one can reconcile this departure from standard interacting theory. In this paper, we provide the answer to this question, determining the generalized relations between the causal Green's functions as a result of the modification of the inner product. Specifically, we find that such approaches are consistent with PHQM.

Here we study PHQM and open and interacting systems on equal field-theoretical footing, aiming to advance a more general and unified description of many-body NH systems. We determine the implications of the pseudo-metric operator on the matrix Green's functions and highlight the key differences and incompatibilities between the various approaches. Our work addresses the ambiguity of ensemble expectation values, providing a comparison of common approaches.  Focusing on fermions in equilibrium, we further develop non-Hermitian linear response theory, which we use to compare the optical conductivity for the (1+1)-dimensional ``Tachyonic'' NH Dirac model, which so far has been studied only within the context of postselection~\cite{Lee2015, Sticlet2022, Sticlet2024}.

\section{Observables and Green's functions} 

\subsection{Isospectral Equivalence}

Pseudo-Hermitian Hamiltonians (satisfying $H^{\dagger} = \eta H \eta^{-1}$) are Hermitian with respect to an inner product modified by a linear, invertible, and Hermitian operator $\eta$ (the pseudo-metric operator). For Hamiltonians with a complete set of biorthonormal eigenvectors and a discrete spectrum, positive definiteness of $\eta$ is the necessary and sufficient condition for $H$ to possess real eigenvalues~\cite{Mostafazadeh2010, Mostafazadeh2002b}. In which case, $H$ is related to an isospectral Hermitian Hamiltonian $h$ by $h = \eta^{1/2}H\eta^{-1/2}$. All diagonalizable $\mathcal{PT}$-symmetric Hamiltonians with a discrete spectrum are pseudo-Hermitian~\cite{Mostafazadeh2002}. In PHQM, the inner product on the Hilbert space is endowed with $\eta$, according to $\langle \cdot|\cdot \rangle_\eta \equiv \langle \cdot|\eta \cdot \rangle$ \footnote{Although this approach can be generalized to an arbitrary metric $\hat{g}$, as done in BQM~\cite{Brody2013}, we focus on the pseudo-metric operator defined by the NHH for its role in recovering unitary time evolution.}. The metric facilitates unitary time evolution and provides an equivalent representation of expectation values of the Hermitian Hamiltonian $h$ in terms of the NHH $H$.

Utilizing the isospectral Hermitian Hamiltonian $h$, real-time observables in the Heisenberg picture $\mathcal{O}(t) \equiv e^{iht}\mathcal{O}e^{-iht}$ are related to observables time-evolved under the NHH via $e^{iht} = \eta^{1/2}e^{iHt}\eta^{-1/2}$. This method ensures manifestly unitary time evolution of the operator $\mathcal{O}(t)$: 
\begin{equation} 
\label{real-time-obs}
\langle \mathcal{O}(t)\rangle = \langle e^{iHt}\widetilde{\mathcal{O}}e^{-iHt}\rangle_{\eta },
\end{equation}
where $\widetilde{\mathcal{O}}\equiv \eta^{-1/2}\mathcal{O}\eta^{1/2}$~\cite{Mostafazadeh2010, Meden2023}. This equation shows the equivalence of the representation of a Hermitian observable $\mathcal{O}$ governed by $h$ in Hermitian QM, with a pseudo-Hermitian observable $\widetilde{\mathcal{O}}$ in a Hilbert space endowed with the positive-definite metric $\eta$ governed by the NHH $H$.

With an interest in applying PHQM to equilibrium many-body systems, we demonstrate this equivalence for an observable computed using the Matsubara formalism. Let $g(k)\equiv (i\omega_n-h)^{-1}$, where $k\equiv (i\omega_n,\bm{k})$, then
\begin{align}
\label{MatsubaraObserv}
\langle \mathcal{O}\rangle &= T \sum_{\omega_n, \bm{k}}\operatorname{tr}  
[\mathcal{O} g(k)]e^{i\omega_n 0^+} \nonumber\\
&= T \sum_{\omega_n, \bm{k}}\operatorname{tr}[\widetilde{\mathcal{O}} G(k)]e^{i\omega_n 0^+}.
\end{align}
The summation is over the fermionic Matsubara frequencies $\omega_n = (2n+1)\pi T$; $T$ is the temperature (units $k_B$) and $n \in \mathds{Z}$. Here we have used $g = \eta^{1/2}G\eta^{-1/2}$ and defined the PHQM Matsubara Green's function $G(k) \equiv (i\omega_n - H)^{-1}$. This definition of the PHQM Matsubara Green's function allows for the calculation of quantities in a similar fashion to Hermitian systems. 

Equivalence between representations is enforced when observables are defined by the isospectral transformation. Since the real-time evolution of the observable in Eq.~\eqref{real-time-obs} is defined strictly in terms of $H$, we can safely obtain $G_{R/A}$ by analytically continuing the PHQM Matsubara Green's function $G(i\omega_n\rightarrow \omega \pm i 0^+ )$. Consequently, the bare retarded and advanced Green's functions for PHQM are $G_{R/A}(\omega) = (\omega - H \pm i 0^+)^{-1}$. In what follows, we will contrast this with the standard prescription for the causal Green's functions when the NH quasiparticle Hamiltonian arises due to the self-energy rather than from the isospectral transformation.

\subsection{Generalized Green's Function Identities}  

Both open and interacting systems can be described in terms of the retarded, advanced, and Keldysh Green's functions $G_{R/A/K}$ respectively. In standard interacting theory, $G_K$ is anti-Hermitian and the causal Green's functions are related by Hermitian conjugation: $G_A = (G_{R})^{\dagger}$. As will be shown, this relation between the causal Green's functions must be modified in PHQM. The bare advanced Green's function for PHQM is defined as $G_A = (\omega - H -i0^+)^{-1}$, hence $(G_A)^{\dagger} = (\omega - H^{\dagger}+i0^+)^{-1}\neq G_R$. Instead, we find that the bare causal Green's functions for PHQM are related by \footnote{We emphasize that this relation holds for the bare Green's functions of PHQM and instances where the retarded and advanced self-energies are related by $\Sigma_{R}=\eta^{-1}\Sigma_A^{\dagger}\eta$.}:
\begin{equation}
G_R = \eta^{-1}(G_{A})^{\dagger}\eta. 
\label{PHQM_GR_GA}
\end{equation}
Despite being described by the same quasiparticle Hamiltonian, PHQM and standard interacting theory offer different prescriptions for the advanced Green's function. For the Hermitian case $H=H^\dagger$, hence $\eta = \mathds{1}$ and thus Eq.~\eqref{PHQM_GR_GA} reduces to the usual relation between $G_{R/A}$. 

The inequivalence between the dynamics of NH systems described by PHQM and standard interacting theory can also be seen in their respective Matsubara Green's functions. To illustrate this point, we compute the Matsubara Green's function for a dissipative interacting system with a frequency-independent quasiparticle Hamiltonian $H = h_0 + i \Gamma$ ($h_0$ and $\Gamma$ are matrix-valued and Hermitian), with a complex spectrum. For a dissipative interacting system, $\Gamma$ is interpreted as the anti-Hermitian part of the self-energy, $\Sigma_R^{\prime \prime}$, and $h_{0}$ contains the real part of the self-energy, $\Sigma_{R}^{\prime}$. Here we take the self-energy to be frequency independent, which is a common Markovian assumption for NH systems and OQSs resulting from time-local self-energies on the Keldysh contour~\cite{Talkington2024, Cayao2022, Groenendijk2021, Kozii2024, Hirsbrunner2019, Chen2018, Bergholtz2019}. Such a self energy is characteristic of elastic scattering and commonly occurs in many-body systems with disorder or impurities. The Matsubara Green's function is defined as $G(i\omega_n) = (2\pi)^{-1}\int d\omega A(\omega)/(i\omega_n-\omega)$, where $A(\omega)$ is the matrix spectral function: $A(\omega) = i(G_R-G_A)$. By evaluating an appropriate contour integral, the Matsubara Green's function evaluates to
\begin{equation}
G(i \omega_n) = \left(i\omega_n - h_0 - i  \Gamma \text{sgn}(\omega_n) \right)^{-1}, 
\label{eq:MatsubaraGF}
\end{equation}
demonstrating that the anti-Hermitian part of the quasiparticle Hamiltonian is accompanied by a sgn$(\omega_n)$ factor in the Matsubara Green's function (as noted in Refs.~\cite{Hirsbrunner2019, Groenendijk2021}). This can be contrasted with the bare Matsubara Green's function for PHQM, which is defined strictly in terms of the NHH: $G(i\omega_n) = (i \omega_n - H)^{-1}$.

It is evident that the Matsubara Green's function for PHQM is incompatible with that of standard interacting systems, since the anti-Hermitian part of $H$ is not accompanied by $\text{sgn}(\omega_n)$. The above result shows that NH systems~\cite{Bitan1, Yu2024, Liu2025} where Matsubara Green's functions do not incorporate a signum function are consistent with a quantum-metric-based approach, namely PHQM. Regardless of the retarded and advanced Green's functions for PHQM being unrelated by Hermitian conjugation, as long as the bare NHH's eigenvalues are real (implying $\eta >0$, defining a positive-definite inner product), the causal Green's functions respect the appropriate analyticity properties, namely the poles of $G_{R/A}$ are in the lower/upper half of the complex plane, respectively. 

PHQM may be used in conjunction with a dissipative interaction. If a system described by PHQM experiences a genuinely dissipative interaction, such as scattering leading to a finite uniform lifetime $\gamma^{-1}$ ($\gamma \geq 0$), then the Matsubara Green's function will pick up a sgn$(\omega_n)$ factor, but only accompanying the $\gamma$ term: $G(i\omega_n) = (i\omega_n - H + i \text{sgn}(\omega_n)\gamma)^{-1}$. The NHH $H$ remains unaccompanied by sgn$(\omega_n)$. Equation~\eqref{PHQM_GR_GA} still holds because on the real frequency axis such a term becomes $i\gamma$.

\subsection{The Role of the Pseudo-Metric Operator} 

The inequivalence between the advanced propagators in standard interacting theory and PHQM can be understood through the effects of equipping the inner product with the pseudo-metric operator. For PHQM with time-independent NHHs, states evolve in time according to $|\psi(t)\rangle = e^{-iHt}|\psi\rangle $. Taking the Hermitian conjugate, we obtain $\langle \psi (t)| = \langle \psi |e^{iH^\dagger t}$. Thus, in the Schr\"{o}dinger picture, time-dependent expectation values take the form $\langle \mathcal{O}(t) \rangle_{\eta} = \langle  \psi|e^{iH^\dagger t}\eta \mathcal{O} e^{-iHt}|\psi\rangle $. To move to the Heisenberg picture, we commute $e^{i H^\dagger t}$ past $\eta$ using the intertwining relation $\eta H = H^\dagger \eta$, to obtain
\begin{equation}
\langle \psi|e^{iH^\dagger t}\eta \mathcal{O} e^{-iHt}|\psi\rangle  = \langle \psi|\eta e^{iH t}\mathcal{O} e^{-iHt}|\psi\rangle.
\end{equation}
Although the ket and bra states are time-evolved by $U(t)$ and $U^\dagger(t)$ respectively, we recover unitary dynamics in the Heisenberg picture, since operators are time-evolved as $U^{-1}(t)\mathcal{O}U(t)$. 

Unlike an OQS, PHQM results in a time evolution of the density matrix given by $\rho(t) = e^{- i H t}\rho e^{i Ht}$, consistent with the von Neumann equation. In contrast, for systems with dissipative interactions, the time evolution of the system degrees of freedom is no longer unitary, nor is it described by the von-Neumann equation. As is exemplified in Markovian OQSs described by the Lindblad master equation, for short times $\dot{\rho} = - i (H\rho - \rho H^{\dagger}) + \cdots$ (where we have left out the recycling term describing quantum jumps, responsible for maintaining the norm of the density matrix). The leading terms are consistent with density-matrix time evolution in the form $\rho(t) \sim e^{-i H t}\rho e^{i H^{\dagger} t}$. The quasiparticle Hamiltonians appearing in $G_R$ and $G_{A}$ can be associated with $H$ and $H^\dagger$ appearing in the forwards $e^{- i H t}$ and backwards $e^{i H^{\dagger}t}$ propagation respectively. This can be contrasted with the von-Neumann equation of motion, which depends on $H$ alone, leading to retarded and advanced propagators that depend solely on $H$.

Despite the concomitant use of PHQM and BQM in open quantum and many-body physics~\cite{Groenendijk2021, Herviou2019, Sim2024, Sun2021, Mandal2010}, we find it to be distinct from the traditional formulation of interacting and open quantum systems. The use of the quantum metric results in a discrepancy between the advanced Green's function defined by interacting and open quantum systems and one consistent with PHQM. Consequently, within PHQM, the non-Hermiticity of the Hamiltonian cannot be motivated from a non-Hermitian self-energy. In the context of PHQM, we can interpret the NH system through its isospectral Hermitian counterpart (related by $h = \eta^{1/2}H\eta^{-1/2}$). If we compare the effect of introducing the non-Hermiticity for the NHH: $h_0\rightarrow h_0+i\Gamma \equiv H$, with the effect of introducing the non-Hermiticity for the isospectral Hermitian Hamiltonian $h_0\rightarrow h$, then we find that the isospectral Hermitian Hamiltonian is renormalized by the difference $h-h_0$. Since both $h$ and $h_0$ are Hermitian, the source of the non-Hermiticity $i \Gamma$ in $H$ arises from the Hermitian (coherent) renormalization of $h_0$ -- \emph{not} a NH self-energy.

\subsection{Implication for Effective Theories}

The previously described constraints have direct implications for NH systems described by effective actions. For concreteness, consider an equilibrium fermionic system with dissipative interactions that have been integrated out, such that the action is quadratic in the fermionic fields $\overline{\psi}, \psi$. In Matsubara space, the action takes the form $S_{\text{eff}}=T\sum_{\omega_n}\overline{\psi}_n\left[-G^{-1}(i\omega_n)\right]\psi_n,$ where $G^{-1}(i\omega_n) = i\omega_n - h_{0} - \Sigma(i\omega_{n})$. A NH quasiparticle Hamiltonian $H = H^\prime+iH^{\prime\prime}$ may be defined by $H^{\prime} = h_0+\Sigma_R^\prime$, $H^{\prime \prime} = \Sigma_R^{\prime \prime}$. As we have previously shown, there are constraints on the form of the frequency dependence of $\Sigma(i\omega_n)$. For concreteness, take $\Sigma_R$ to be frequency independent and the decay rates to be described by the retarded self-energy $\Sigma_R = \Sigma^\prime_R+i\Sigma_R^{\prime \prime}$, as before. Consequently, the Mastsubara self-energy must have the frequency dependence $\Sigma(i\omega_n) = \Sigma^\prime_R+i\text{sgn}(\omega_n)\Sigma_{R}^{\prime \prime}$, as in Eq.~\eqref{eq:MatsubaraGF}. In imaginary time, the effective action reads
\begin{align}
S_{\text{eff}} &=\int_{0}^{1/T}\overline{\psi} (\tau)\big[\left(\partial_\tau + H^{\prime}\right)\delta(\tau-\tau^\prime) \nonumber\\& \quad\quad + 
T\text{csc}\left(\pi T(\tau-\tau^{\prime})\right)H^{\prime \prime}\big]\psi(\tau^{\prime})d\tau d\tau^{\prime},
\label{action}
\end{align}
where we have computed the Fourier transform of the frequency-space action defined in terms of the inverse Green's function $G^{-1}(i\omega_n) = i\omega_n - H^\prime -i \text{sgn}(\omega_n)H^{\prime \prime}$. Although the NH term in the action is translationally invariant, it is non-local in imaginary time. 

We emphasize that NH terms in an effective action for an equilibrium system must have the appropriate frequency (or time) dependence in Matsubara space to yield causal Green's functions satisfying the constraint $G_R=G_A^\dagger$. Many works can be found studying NH systems with an effective action (e.g., certain NH BCS theories), which have NH terms without time or frequency dependence, that is, of the form $S_{\text{eff}}\sim \overline{\psi}\partial_\tau \psi+H[\overline{\psi}, \psi]$, where $H$ is time independent. Although such systems lack compatibility with standard interacting physics, some can be reinterpreted through the path-integral formulation of PHQM~\cite{Mostafazadeh2007}. In the next section, we implement these ideas in the context of a specific model.

\section{NH Dirac model}

\subsection{Isospectral Hamiltonian}

Here we consider the Tachyon model, a paradigmatic (1+1)-dimensional NH Dirac Hamiltonian characterized by a real mass (gap parameter) $\Delta$ and an imaginary mass $m$. The Hamiltonian is $H=v_F k \sigma_x+\Delta \sigma_y-i m \sigma_z-\mu \mathds{1}$, and in terms of the second-quantized fermionic operators:
\begin{equation}
\begin{aligned}
\hat{H} & =\sum_k\left(\begin{array}{cc}
c^{\dagger}_{k \uparrow} & c^{\dagger}_{k \downarrow}
\end{array}\right)\left(\begin{array}{cc}
-\mu-i m & v_F k-i \Delta \\
v_F k+i \Delta & -\mu+i m
\end{array}\right)\binom{c_{k \uparrow}}{c_{k \downarrow}}. \\
\end{aligned}
\end{equation}
The eigenvalues are $\pm\sqrt{v_F^2 k^2+\Delta^2-m^2}-\mu$. The effective gap $\Delta_e^2  \equiv \Delta^2-m^2 $ demarcates the various parameter regimes of the model: $\Delta_e^2>0$ insulating ``gapped phase'', $\Delta_e^2=0 $ Dirac-like ``linear phase'', and $\Delta_e^2<0 $ hyperbolic ``Tachyon'' phase. When $v_F^2 k^2+\Delta^2-m^2 = 0$, the system exhibits exceptional points, where both the eigenvectors and eigenvalues coalesce. 

Taking the Tachyon model to describe an interacting or open-quantum system, we require $\Sigma_{R}^{\prime\prime}$ to be negative definite to ensure dynamical stability \footnote{Dynamical stability requires the anti-Hermitian part of the dynamical matrix (the effective NHH) to be negative definite. For Fermionic OQSs described by quadratic Lindbladians, $\Sigma_R^{''}$ is negative definite, guaranteeing dynamical stability~\cite{Thompson2023}.}. Here we will achieve this by adding a uniform lifetime to the quasiparticle Hamiltonian $H\rightarrow H -i \gamma \mathds{1}$, with the decay rate $\gamma$, where $\gamma > |m| $ to ensure that $\Sigma_{R}^{\prime\prime}$ is negative definite (while $\Delta/m$ remains arbitrary). Although $\gamma$ describes the uniform decay rate for the system, $m$ may be thought of as a relative decay rate between spins with inverse lifetimes $\tau^{-1}_{\uparrow/\downarrow} = 2(\gamma \pm m)$.

We also consider the NH Dirac model described by PHQM. Here we are not constrained to require the anti-Hermitian part of the Hamiltonian, $\Gamma$, to be negative definite since $\Gamma$'s non-Hermiticity does not describe dissipation, rather a Hermitian renormalization of an isospectral Hermitian system. Nevertheless, for the purpose of comparing the same quasiparticle Hamiltonian against different approaches, we consider the pseudo-Hermitian system with a constant decay rate $\gamma$. However, we are not restricted to $\gamma>|m|$, since PHQM describes an isospectral Hermitian system. While considering the model subject to PHQM, we restrict our analysis to the case of $\Delta^2 > m^2$ to ensure a positive-definite metric operator. 

By evaluating $\eta$ in terms of the right eigenstates of $H^\dagger$ and taking its square root, we obtain the isospectral Hermitian Hamiltonian:
\begin{equation} \label{isospectralHamiltonian}
h = \sqrt{\frac{v_F^2k^2+\Delta^2-m^2}{v_F^2 k^2+\Delta^2}}(v_F k \sigma_x + \Delta \sigma_y) - \mu \mathds{1}.
\end{equation}
The pseudo-Hermitian system may be mapped to the isospectral Hermitian system provided $\Delta_e > 0$. The ``tachyonic'' NHH is mapped to a Dirac Hamiltonian with momentum-dependent energy scale renormalization. In the IR limit, the energy scale is suppressed for $m\lesssim \Delta$, with a velocity and mass renormalization $\approx \sqrt{1-(m/\Delta)^{2}}$. In the UV limit, the isospectral system becomes a free Dirac fermion.

\subsection{Gauge Invariance and Current Operators}

To derive the electromagnetic response, we apply minimal coupling to the Hamiltonian: $k \rightarrow k -e  A$. From the general definition~\cite{DupuisBook} of the local current operator, applied to a noninteracting system the current can be written as: $J = -e\partial_k G^{-1}(k)=e\partial_{k}H(k)$. We also consider the transformed current $\widetilde{j} \equiv \;\eta^{-1/2}j \eta^{1/2}$ with $j = e \partial_k h(k)$, defined by the isospectral Hermitian Hamiltonian $h$. For some NHHs $J = \widetilde{j}$, but in general the two currents are inequivalent. In particular, $J = \widetilde{j} + e\big[H,\widetilde{\partial}_k\big]$, where $\widetilde{\partial}_k \equiv \eta^{-1/2}\partial_{k} \eta^{1/2}$. For the NH Dirac model studied here, the commutator does not vanish. The current operators are $J = e \boldsymbol{v}(\boldsymbol{k})$ and $\widetilde{j} = e \widetilde{\boldsymbol{v}}(\boldsymbol{k})$, where the velocity operators are given in Appendix~\ref{App:IsospecMatsubara}.

The inequivalence of the two current operators originates from the possible dependence of the pseudo-metric operator on $A$. Gauge invariance requires the NHH to couple to the vector potential $H=H[A]$. Typically, the pseudo-metric operator is defined in terms of the eigenstates of the Hamiltonian, leading to momentum dependence and therefore vector potential dependence in $\eta$. For the NHH to be observable, it must maintain its intertwining relation ($\eta H=H^{\dagger}\eta$) with the pseudo-metric operator. Although equivalent electromagnetic response with the isospectral Hermitian system can be obtained by forgoing the vector potential dependence of the pseudo-metric operator, it comes at the cost of observability of $H[A]$. To summarize, observability of the gauge-invariant system dictates a preferred electromagnetic current (either that of the NHH or its isospectral counterpart) when the pseudo-metric operator is momentum-dependent.

\section{Linear-response and Electrical Conductivity} 

\subsection{Kubo Formula}

Here we generalize the standard Kubo formula for noninteracting systems to NH systems described by PHQM. For a system that satisfies the fluctuation-dissipation theorem (FDT), we may utilize the Matsubara formalism and evaluate the linear response for frequency-independent operators $A$, $B$: $\chi_{AB}(q) =  \sum_{\omega_n, \bm{k}} \operatorname{tr}[G(k) A G(k+q) B],$ where $q\equiv(i\Omega_m,\bm{q})$ for the bosonic Matsubara frequencies $\Omega_m = 2\pi m T;\; m\in \mathds{Z}$. For a system described by PHQM, where the source of the non-Hermiticity follows from the Hermitian renormalization $h_0\rightarrow h$ (rather than from the dissipative renormalization by $\Sigma^{\prime \prime}_{R}$), we can evaluate the response using the Green's functions of PHQM~\eqref{MatsubaraObserv}. See Appendix~\ref{PHQMConductivity1}.

We compute optical conductivity~\cite{Abrikosov1963, Antezza2025} using standard finite-temperature field-theoretic approaches, considering non-Hermiticity treated as a dissipative interaction and as a nondissipative interaction consistent with unitary dynamics. For the former case, the NH terms in the NHH can be identified with the self-energy. In the latter case, we recover unitary dynamics by using PHQM. In each instance, we consider the same quasiparticle Hamiltonian, albeit subject to a different physical approach. In all cases, we take the system to be in thermal equilibrium, where the FDT holds. Using PHQM, we consider two definitions for the current operator: (i) $J$, derived from the NHH $H$, and (ii) $\widetilde{j}$ derived from the isospectral Hermitian Hamiltonian $h$. The significance of considering the first definition for the current is that it is representative of NHHs treated with unitary dynamics (which is consistent with the approach many studies have adopted for NH systems~\cite{Meden2022, Bitan1, Murshed2025, Bitan3, Juricic2025, Liu2025}), and the significance of the second definition is that the NHH may be directly connected with the isospectral Hermitian system.

\begin{table}[h] 
\centering
\begin{tabular}{c} 
\hline
\multicolumn{1}{c} {\textbf{DC Conductivity for Tachyon Dirac Model} } \\
\hline
\hline  Postselection~\cite{Sticlet2022}:  \\
\\[-.8em]
$\dfrac{\pi}{2} \dfrac{\sqrt{m^2-\Delta^2}}{m^2}\; \theta \left(m^2-\Delta^2\right)$  \\ 
\\[-.8em]
\hline  Standard interacting physics ($\Sigma_R^{\prime\prime}=H^{\prime\prime}$): \\
\\[-.8em]
$\dfrac{\gamma^{2}-m^{2}}{(\gamma^2+\Delta^2-m^2)^{3/2}}$ \\
\\[-.8em]
\hline  PHQM with $J$: \\
\\[-.8em]
$\dfrac{\gamma^{2}}{(\gamma^2+\Delta^2-m^2)^{3/2}}$\\ 
\\[-.8em]
\hline  PHQM with $\widetilde{j}$: \\
\\[-1em]
$\begin{cases}
& \gamma^{-1} + \left(2 |\Delta| - \dfrac{m^2}{|\Delta|}-2\sqrt{\Delta^2-m^2}\right)\gamma^{-2}+\mathcal{O}(\gamma^{-3}) \\
& \left(\dfrac{8\Delta^4 - 8 \Delta^2 m^2 + m^4}{4m^4(\Delta^2-m^2)^{3/2}}-\dfrac{m^2+2\Delta^2}{m^4|\Delta|}\right)\gamma^2 + \mathcal{O}(\gamma^4) \\
\end{cases}$\\
\\[-.8em]
\hline
\end{tabular}
\caption{$\sigma_{\DC}$ $[e^2 v_F/(2 \pi)]$ for the Tachyon Model evaluated using various approaches at $T=0$, $\mu=0$. $e$ is the electronic charge, $v_F$ the Fermi velocity, $\Delta$ the real mass, $m$ the imaginary/NH mass, and $\gamma$ the uniform decay rate. The DC conductivity using PHQM with $\widetilde{j}$ is computed for $\gamma\gg1$ and $\gamma\ll1$.}
\label{table1}
\end{table}

\begin{widetext}
\begin{figure*}[t]
\centering\includegraphics[width=1\linewidth]{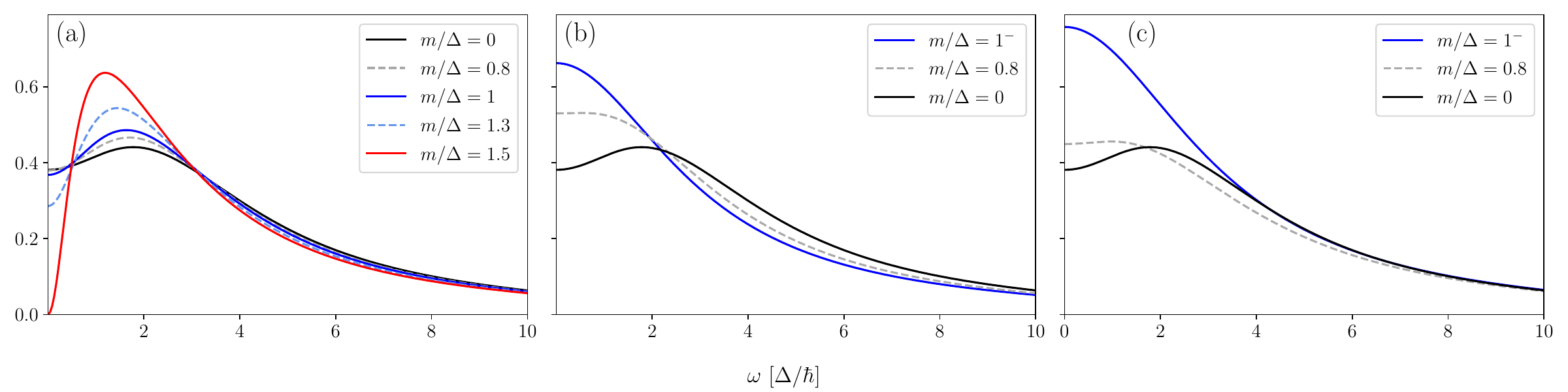}
\caption{Real part of the zero-temperature electrical conductivity $\sigma^\prime(\omega)$ (units $e^2v_F/(2\pi\Delta)$; $\gamma/\Delta =1.5$), for several approaches.  (a) Standard (dissipative dynamics) with $J$, (b) PHQM (unitary dynamics) with $J$, (c) PHQM (unitary dynamics) with $\widetilde{j}$.}
\label{Fig:OpticalConductivity}
\end{figure*}
\end{widetext}

The electrical conductivities are shown in Fig.~\ref{Fig:OpticalConductivity}, and their DC limits are given in Table~\ref{table1}. The optical conductivity of the non-Hermitian Tachyon model was previously studied in Ref.~\cite{Sticlet2022}, in a context where its dynamics are consistent with postselection. Their approach (which has been generalized to finite temperature~\cite{Myself_unpublished}), is included in Table~\ref{table1} for comparison~\footnote{We have adjusted their parameters and units to match our own.}. All conductivities shown in Fig.~\ref{Fig:OpticalConductivity} satisfy their respective optical sum rule (OSR), which is determined from the Kramers-Kr\"{o}nig relations of the respective response formulae. Approaches (a),(b) (using the current operator $J$) obey the same optical sum rule, namely $\int_{-\infty}^{\infty}\sigma^{\prime}(\Omega)d\Omega = e^2v_F$, whereas for approach (c), the isospectral Hermitian system has an unconventional sum rule. In the dissipationless limit ($\gamma\rightarrow 0^+$), we find that the sum rule is $\int_{-\infty}^{\infty}\sigma^{\prime}(\Omega)\ d\Omega =e^2v_F \left[1 + \left(\bar{m}^{-1}-\bar{m}\right) \text{artanh}(\bar{m})\right]/2,$ with the dimensionless parameter $\bar{m}\equiv m/\Delta$ characterizing the strength of the non-Hermiticity. As required, the generalized OSR recovers the conventional result in the limit $\overline{m}\rightarrow 0$. The generalization for finite $\gamma$ is computed in Appendix~\ref{App:IsospecMatsubara}.

In the limit $m\rightarrow0$, all three non-postselected approaches recover the same result: $\sigma_{\DC} = \frac{e^2v_F}{2\pi}\frac{\gamma^2}{(\gamma^2+\Delta^2)^{3/2}}$, describing a massive (1+1)-dimensional Dirac fermion with decay rate $\gamma$. For the standard approach, the imaginary mass $m$ is taken to be a NH self-energy, resulting in the renormalized scattering rate $\gamma \rightarrow \sqrt{\gamma^2 -m^2}$. This can be contrasted with the use of PHQM with current $J$, which results in renormalization of the real mass $\Delta \rightarrow \sqrt{\Delta^2-m^2}$. For the case of PHQM with current $\widetilde{j}$, the renormalization of the real mass $\Delta$ is less simple, as it does not appear in quadrature with $im$. This is expected due to the non-trivial energy-scale renormalization appearing in Eq.~\eqref{isospectralHamiltonian}.

The renormalization of the mass instead of the quasiparticle lifetime highlights the difference between the two approaches: in the standard interacting approach, the NH mass is a relative inverse lifetime, describing dissipative effects while within PHQM, the NH mass is an actual mass, describing coherent non-dissipative behavior. Consequently, the impact of the non-Hermiticity on the transport for each approach differs, either enhancing or diminishing the DC conductivity. Recall that the source for the non-Hermiticity in NHHs described by PHQM does not come from the anti-Hermitian part of the self-energy. We can interpret a NHH $H$ described by PHQM through the isospectrally related Hermitian Hamiltonian $h = h_0 + \Sigma_{R}^{\prime (h)}$, which is renormalized by a \emph{Hermitian} self-energy $\Sigma_{R}^{\prime (h)} = (h-h_0)$. In PHQM the non-Hermiticity of the NHH describes the Hermitian renormalization of an isospectrally equivalent Hamiltonian via $\Sigma^{\prime (h)}_R$, whereas in standard interacting or open quantum systems the non-Hermiticity describes dissipation introduced by a NH self-energy.\vspace{2mm} 


\subsection{Equilibrium Expectation Values}

As we emphasize in this paper, the origin of the non-Hermiticity results in different prescriptions for the matrix Green's functions. Furthermore, for systems in thermodynamic equilibrium, the distribution is related to the causal Green's functions by the FDT. Thus, the distinctions in the Green's functions are necessary for evaluating thermodynamic averages of equilibrium systems. To clarify this point, we display the respective zero-temperature expectation values for (non-degenerate) systems described by the approaches considered here:
\begin{widetext}
\begin{equation} \label{variousEVs}
  \langle \mathcal{O}\rangle =
    \begin{cases}
      \frac{i}{2 \pi}\sum_{n} \left[\text{log}(\xi_n)\langle L_n|\mathcal{O}|R_n\rangle - \text{log}(\xi_n^*)\langle R_n|\mathcal{O}|L_n\rangle\right] & \text{(i) Dissipative Dynamics in thermal equilibrium \citep{Shen2024}}\\
     -\frac{1}{\pi}\sum_{n} \text{arg}(\xi_n)\langle L_n|\mathcal{O}|R_n\rangle \xrightarrow{\text{Im($\xi_n$)}\rightarrow 0} \langle L_0|\mathcal{O}|R_0\rangle & \text{(ii) PHQM in thermal equilibrium } \\
     \langle R_0|\mathcal{O}|R_0\rangle & \text{(iii) Postselected Dynamics stationary state}
    \end{cases}       
\end{equation}
\end{widetext}

\clearpage 
\newpage 
\vspace{5mm}

The different expressions here correspond to dissipative dynamics in thermal equilibrium~\cite{Shen2024}, PHQM in thermal equilibrium, and postselected dynamics in a stationary state, respectively. The summation in $n$ is over the eigenstates of the system. For real eigenvalues, the groundstate (denoted by $n=0$) is defined as the state with lowest energy, while in the case of complex eigenvalues under postselected dynamics, $|R_0\rangle$ is instead the exponentially preferred state with the maximum imaginary part \cite{Kou2022}. The first line of Eq.~\eqref{variousEVs} shows the form of the expectation value evaluated in the zero-temperature limit for a dissipative system in thermal equilibrium, which was computed in Ref.~\citep{Shen2024}. In the second line, we extend this result to systems described by PHQM, and evaluate the case of real eigenvalues (long-lived quasiparticles). The third line of Eq.~\eqref{variousEVs} is the zero-temperature limit of the thermal NH (stationary state) under postselected dissipative dynamics. We find that the right-right and left-right expectation value is appropriate for postselected dynamics and PHQM respectively, whereas for standard (not employing a quantum-metric) equilibrium interacting systems both the right-left and left-right arise.

\section{Conclusion}

NHHs have grown in popularity as an outlet for exploring exotic aspects of OQSs and dissipative interactions. In contrast to common usage of the biorthogonal inner product in evaluating ground-state expectation values or time evolution with the von-Neumann equation using a NHH, we demonstrate that these approaches are inconsistent with dissipative dynamics. We show that both instances are consistent with PHQM, described by unitary time evolution with a NHH. Such approaches are implicitly consistent with an inner product modified by a pseudo-metric operator (even when it is not overtly used~\cite{Rivers2011, Mostafazadeh2007}). 

We showed that the implicit use of the pseudo-metric operator in systems described by PHQM results in a generalized relation between the causal Green's functions, namely $G_R = \eta^{-1}G_A^{\dagger}\eta$. Matsubara Green's functions without the signum function $\textrm{sgn}(\omega_{n})$ accompanying the anti-Hermitian part of the quasiparticle Hamiltonian are inconsistent with standard interacting theory. Instead, such approaches are compatible with the use of a quantum-metric, described by PHQM. This distinction is significant, as PHQM describes unitary time evolution, whereas the anti-Hermitian part of the self-energy arises due to non-unitary effects from dissipative dynamics.

The different results for the electrical conductivity and equilibrium expectation values serve as experimental signatures to distinguish the various approaches to non-Hermitian physics that can be associated with particular physical phenomena.

\section{Acknowledgments}

We thank Joshuah Heath and Nikolay Gnezdilov for insightful conversations. This research was supported through funding provided by Dartmouth College.

\appendix

\section{Motivation}

\setcounter{subsection}{0}
\subsection{Non-Hermiticity in Open Quantum Systems} 

Consider a system with appropriate separation of energy scales from the environment and weak system-environment coupling. These assumptions ensure that the Born-Markov approximations are applicable. As a result, dissipation to an environment can be described by the Gorini-Kossakowski-Sudarshan-Lindblad (GKSL)  master equation~\citep{Lindblad1976, Gorini1976, Manzano2020, BreuerBook}:
\begin{align}
\dot{\rho} &= - i \left[H,\rho \right] + \sum_{j}\gamma_{j}\left(L_{j}\rho L_{j}^\dagger - \frac{1}{2}\{L^\dagger_{j} L_{j},  \rho \} \right) \nonumber\\
&= - i (H_{\text{cond}}\rho - \rho H^{\dagger}_{\text{cond}}) + \sum_{j} \gamma_j  L_j \rho L_j^\dagger.
\label{eq:GKSLEqn}
\end{align}
$H$ is a Hermitian Hamiltonian, $\{L_j\}$ are the Lindblad jump operators which describe dissipation to the environment, and $\gamma_j$ are the respective decay rates; here $\hbar=1$. Although the commutator in the first line of Eq.~\eqref{eq:GKSLEqn} describes unitary time evolution, the second and third terms describe dissipative non-unitary time evolution. The third term (with jump operators acting on both sides of the density matrix), often called the \emph{recycling} term, is responsible for quantum jumps. In the second line of Eq.~\eqref{eq:GKSLEqn}, we have defined a conditional non-Hermitian Hamiltonian (NHH)~\citep{Meden2023}:
\begin{equation}
H_{\text{cond}} \equiv H -  \frac{i}{2}\sum_{j} \gamma_jL^\dagger_j L_j.
\end{equation}

In the absence of quantum jumps, the recycling term can be discarded. The GKSL master equation then reduces to a generalized von-Neumann equation with a conditional NHH:
\begin{equation}
\dot{\rho} = - i (H_{\text{cond}}\rho - \rho H^{\dagger}_{\text{cond}}).
\label{generalizedVN}
\end{equation}
This approach provides an approximate method for describing dissipative dynamics for short time scales or, alternatively, an exact method, referred to as \emph{postselection}, conditioned on the absence of quantum jumps. To enforce normalization of $\rho$, it is customary to redefine the time evolution by $\rho(t) \rightarrow \rho(t)/\mathrm{Tr}[\rho(t)]$, which leads to an effective non-linear master equation describing the dynamics conditioned on the absence of quantum jumps. Dynamics of this form have appeared in other contexts using postselection, such as in the ancilla approach~\cite{Meden2023}. Despite the conditional Hamiltonian being one of the primary outlets for considering non-Hermiticity from within an open-quantum-systems context, without stating the gain-loss structure of the system, we have not fully specified the open quantum system. At the level of the Lindblad Master equation, we can express the jump operators in terms of the gain $\Gamma^{(G)}$ and loss $\Gamma^{(L)}$ matrices which describe the pumping and decay of particles, respectively:
\begin{align}\nonumber
\dot{\rho} &= - i (H_{\text{cond}}\rho - \rho H^{\dagger}_{\text{cond}})\\ &\quad\quad+ \sum_{j l} \Gamma^{(G)}_{j l} c^{\dagger}_j \rho c_l +  \sum_{j l} \Gamma^{(L)}_{j l} c_j \rho c^{\dagger}_l,  
\end{align}
where $H_\text{cond} \equiv H - \frac{i}{2}(\Gamma^{(L)} - \Gamma^{(G)})$. As highlighted in Ref.~\citep{Clerk2022}, it is possible to take into account the full unconditional dynamics by instead considering an effective Hamiltonian of the form $H_{\text{eff}} \equiv  H - \frac{i}{2}(\Gamma^{(L)} + \Gamma^{(G)})$, while also incorporating the effect that the gain and loss have on the distribution function for the system. Formally, Lindbladian dynamics can be encapsulated in terms of the Keldysh action within a closed-time contour integral~\cite{Kamenev_2023, Talkington2024, Thompson2023, GomezLeon2022}.

\subsection{Non-Hermiticity via the Self Energy} 

A method of obtaining NHHs, more customary in the condensed matter physics community~\citep{Kozii2024, 
Groenendijk2021, Hirsbrunner2019, Chen2018}, is to define the quasiparticle Hamiltonian $H$ appearing in the dressed retarded Green's function:
\begin{equation}
G_{R}(\omega) = (\omega - H)^{-1}, \quad H \equiv H_0 + \Sigma_{R}(\omega).
\end{equation}
The Hermitian single-particle Hamiltonian is $H_0$ and the (retarded) self-energy is $\Sigma_{R} = \Sigma_R^{\prime}+ i \Sigma_R^{\prime \prime}$. The NHH $H$ is typically assumed to be frequency independent~\citep{Cayao2022, Bergholtz2019, Groenendijk2021, Kozii2024, Hirsbrunner2019, Chen2018}. For NHHs motivated from the interaction self energy, this may be interpreted as a wide-band limit $\Sigma_{R}(\omega)|_{\omega=0}$~\citep{Stefanucci2013NonequilibriumMT}. Frequency dependence in the self energy can describe non-Markovian effects of the interaction~\citep{Michishita2020}.

\section{Second quantization of NH systems}

Consider a finite-dimensional non-degenerate NHH $H$, with right $\left|R_{\alpha}\right\rangle$ and left $\left|L_{\alpha}\right\rangle$ eigenstates indexed by $\alpha\in\mathbb{N}$. The eigenvector relations are given as follows:
\begin{equation} 
\begin{aligned}
H\left|R_{\alpha}\right\rangle=\xi_{\alpha}\left|R_{\alpha}\right\rangle \quad &\Rightarrow \quad \left\langle R_{\alpha}\right| H^{\dagger}=\xi_{\alpha}^*\left\langle R_{\alpha}\right|\\
H^{\dagger}\left|L_{\alpha}\right\rangle=\xi_{\alpha}^*\left|L_{\alpha}\right\rangle \quad &\Rightarrow \quad \left\langle L_{\alpha}\right| H=\xi_{\alpha}\left\langle L_{\alpha}\right|.
\end{aligned}
\end{equation}
The single-particle eigenstates can be expressed in terms of second-quantized creation and annihilation operators:
\begin{equation}
\left|R/L_{\alpha}\right\rangle = c^{\dagger R/L}_{\alpha}|0\rangle.
\end{equation}
The eigenstates are biorthonormal: $\langle L_{\alpha}|R_\beta\rangle = \delta_{\alpha \beta}$, and the resolution of unity is $\mathds{1} = \sum_{\alpha}|R_{\alpha}\rangle\langle L_{\alpha}|$.

The right and left single-particle energy eigenstates can be expanded in terms of a complete orthonormal basis:  
\begin{equation}
\begin{aligned}
& \left|R_\alpha\right\rangle=\sum_i r_{\alpha i}|i\rangle \\
& \left|L_\alpha\right\rangle=\sum_i l_{\alpha i}|i\rangle
\end{aligned} \Rightarrow \begin{aligned}
& c_\alpha^{\dagger R}=\sum_i r_{\alpha i} c_i^{\dagger} \\
& c_\alpha^{\dagger L}=\sum_i l_{\alpha i} c_i^{\dagger}.
\end{aligned}
\end{equation}
Here we distinguish the single-particle orthonormal states $|i\rangle$ (Latin indices), from the single-particle biorthonormal eigenstates $|R_\alpha\rangle$, $|L_\alpha\rangle$ (Greek indices). The orthonormal states satisfy $\langle i | j\rangle = \delta_{i j}$, and have associated creation and annihilation operators $c_i, c^\dagger_i$ obeying standard Fermi-Dirac statistics. Since $c_i^{\dagger}|0\rangle=|i\rangle$, we can recognize $|0\rangle$ as the vacuum state for both the biorthonormal creation and annihilation operators, as well as the orthonormal creation annihilation operators. 

A particular bilinear of the biorthonormal creation annihilation operators that satisfies Fermi-Dirac statistics is: 
\begin{align}\nonumber
\{ c_\alpha^L , c^{\dagger R}_\beta \} = \sum_{ij}l^{*}_{\alpha i }r_{\beta j }\{c_i,c^\dagger_j\} &= \sum_{ij}l^{*}_{\alpha i }r_{\beta j }\delta_{ij}\\ &= \langle L_\alpha | R_\beta  \rangle = \delta_{\alpha \beta}.
\end{align}
Here we collect all relevant quantum numbers (such as band-index, spin, or momentum) into a single index. In the case of a continuous variable, the Kronecker delta can be expressed as a Dirac delta function. Note that not all combinations of left or right creation annihilation operators recover Fermi-Dirac statistics:
\begin{align}\nonumber
\left\{c_\alpha^L, c_\beta^{\dagger L}\right\} = \sum_{i j} l_{\alpha i}^* l_{\beta j}\{c_i, c_j^{\dagger}\} &= \sum_{i j} l_{\alpha i}^* l_{\beta j} \delta_{i j}\\
&= \langle L_\alpha |L_\beta \rangle \neq \delta_{\alpha \beta}.
\label{finiteoverlap}
\end{align}
A similar relation, $\{c_\alpha^R, c_\beta^{\dagger R}\} = \langle R_\alpha | R_\beta \rangle$, holds for the right creation and annihilation operators. Here, we may take expressions such as $\langle R_\alpha | R_\beta \rangle$ to implicitly include $\delta_{k,k^{\prime}}$ when momentum is part of the index. Although non-Hermiticity leads to overlap between eigenstates, here, all momentum modes are taken to be independent and orthogonal.

In the case of a bilinear operator involving the same kind of biorthonormal operator ($c$ or $c^{\dagger}$), we obtain 
\begin{equation}
\{c_\alpha^{\dagger R} , c^{\dagger R}_\beta \} = \sum_{ij}r_{\alpha i }r_{\beta j }\{c^\dagger_i,c^\dagger_j\} = 0,
\label{anticom1}
\end{equation}
since $\{c^\dagger_i,c^\dagger_j\} = 0$. Similarly,  $\{ c_\alpha^{\dagger L} , c^{\dagger L}_\beta \} = \{ c_\alpha^{ L} , c^{ L}_\beta \} = \{ c_\alpha^{\dagger R} , c^{\dagger R}_\beta \} = \{ c_\alpha^{R} , c^{R}_\beta \} = \{ c_\alpha^{L} , c^{R}_\beta \}= \{ c_\alpha^{\dagger L} , c^{\dagger R}_\beta \}=0$.  
This formalism of biorthonormal quasiparticle creation/annihilation operators is discussed in the Supplemental Material of Ref.~\cite{Meden2023}. \\

\section{Linear-response Formulae}
\setcounter{subsection}{0}
\subsection{Pseudo-Hermitian QM}

For systems described by PHQM with real energy eigenvalues, the Hilbert space is endowed with a positive-definite metric operator $\eta$. Such systems are Hermitian with respect to $\eta$. That is, the condition $\langle{H\psi}|\phi\rangle_{\eta}=\langle\psi|H\phi\rangle_{\eta}$ for all $|\psi\rangle, |\phi\rangle\in\mathcal{H}$ is equivalent to the NHH satisfying $\eta H \eta^{-1} =H^{\dagger}$. Defining the inner product with the pseudo-metric operator results in the recovery of unitary dynamics described by the von-Neumann equation. Thus, the linear response formula takes its usual form:
\begin{equation}
\label{correlation1}
\chi_{A B}\left(t-t^\prime\right) =-i \theta(t-t^\prime)  \langle [\hat{A}(t-t^\prime),\hat{B}(0)] \rangle,
\end{equation}
with the operator time dependence described by the usual Heisenberg picture time evolution: 
\begin{equation}
\hat{A}(t-t^\prime) = e^{i H (t-t^\prime)}\hat{A}e^{-i H (t-t^\prime)}.
\end{equation} 
This may be contrasted with the time evolution used for NH systems within the context of postselection, where the dynamics is of the form $\hat{A}(t) = e^{iH^\dagger t}\hat{A} e^{-iHt}$. To evaluate the response, we decompose the single-particle operators in a biorthogonal single-particle basis:
\begin{align} \nonumber
\hat{A}(t) &=\sum_{\bm{k},\alpha \beta} e^{i(\xi_{\bm{k},\alpha} - \xi_{\bm{k},\beta}) t} \langle L_{\bm{k}, \alpha}| \hat{A} |R_{\bm{k},\beta} \rangle c^{\dagger R}_{\bm{k},\alpha} c^{L}_{\bm{k},\beta},\\ \quad \quad     \hat{B} &= \sum_{\bm{k^\prime},\gamma \delta} \langle L_{\bm{k^\prime},\gamma}| \hat{B} |R_{\bm{k^\prime},\delta }\rangle c^{\dagger R}_{\bm{k^\prime},\gamma} c^{L}_{\bm{k^\prime},\delta}. 
\label{PHQM_AB}
\end{align}
Here, we project onto a left-right basis, so that time dependence may be factored out. Inserting Eq.~\eqref{PHQM_AB} into Eq.~\eqref{correlation1} (suppressing the momentum index), we have:
\begin{widetext}
\begin{equation*}
\chi_{A B}\left(t-t^\prime\right) =-i \theta(t-t^\prime) \sum_{\alpha \beta \gamma \delta} e^{i(\xi_{\alpha} - \xi_{\beta}) (t-t^\prime)} \langle L_{ \alpha}| \hat{A} |R_{\beta} \rangle  \langle L_{\gamma}| \hat{B} |R_{\delta }\rangle   \langle[c^{\dagger R}_{\alpha} c^{L}_{\beta} ,c^{\dagger R}_{\gamma} c^{L}_{\delta}] \rangle.
\end{equation*}
By performing the Fourier transform and evaluating the commutator, we obtain
\begin{align}
\chi_{A B}\left(\Omega\right) &= \sum_{\alpha \beta \gamma \delta}  \frac{\langle L_{\alpha}| \hat{A} |R_{\beta} \rangle  \langle L_{\gamma}| \hat{B} |R_{\delta }\rangle }{\Omega +\xi_{\alpha} - \xi_{\beta} + i 0^{+}}  \langle c^{\dagger R}_{\alpha} c^{L}_{\beta} c^{\dagger R}_{\gamma} c^{L}_{\delta} - c^{\dagger R}_{\gamma} c^{L}_{\delta} c^{\dagger R}_{\alpha} c^{L}_{\beta} \rangle \nonumber\\
&=\sum_{\alpha \beta \gamma \delta}  \frac{\langle L_{\alpha}| \hat{A} |R_{\beta} \rangle  \langle L_{\gamma}| \hat{B} |R_{\delta }\rangle }{\Omega +\xi_{\alpha} - \xi_{\beta} + i 0^{+}}  \langle c^{\dagger R}_{\alpha} (\delta_{\gamma \beta} -c^{\dagger R}_{\gamma}c^{L}_{\beta} ) c^{L}_{\delta} 
- c^{\dagger R}_{\gamma}(\delta_{\alpha \delta} - c^{\dagger R}_{\alpha} c^{L}_{\delta}) c^{L}_{\beta} \rangle \nonumber\\
&=\sum_{\alpha \beta \gamma \delta}  \frac{\langle L_{\alpha}| \hat{A} |R_{\beta} \rangle  \langle L_{\gamma}| \hat{B} |R_{\delta }\rangle }{\Omega +\xi_{\alpha} - \xi_{\beta} + i 0^{+}}   (\delta_{\gamma \beta}\langle c^{\dagger R}_{\alpha} c^{L}_{\delta}\rangle  - \delta_{\alpha \delta} \langle c^{\dagger R}_{\gamma} c^{L}_{\beta}  \rangle).
\end{align}

If we take the density matrix to have the form $\rho = e^{-\beta H}/Z$, (which is stationary under time evolution according to the von Neumann equation), then $\langle c^{\dagger R}_{\alpha} c^{L}_{\beta}  \rangle = \delta_{\alpha \beta} n_F(\xi_\alpha)$. After restoring the momentum dependence, we then obtain the Kubo formula for a noninteracting system described by PHQM: 
\begin{equation}
\chi_{A B}(\Omega)  = \sum_{\bm{k}}\sum_{\alpha \beta} \langle L_{\bm{k},\alpha}| A|R_{\bm{k}+\bm{q},\beta} \rangle\langle L_{\bm{k}+\bm{q},\beta}|B|R_{\bm{k},\alpha} \rangle \frac{n_{F}(\xi_{\bm{k},\alpha}) - n_{F}(\xi_{\bm{k}+\bm{q},\beta})}{\Omega +\xi_{\bm{k},\alpha} - \xi_{\bm{k}+\bm{q},\beta} +i0^+}. 
\label{PHQM Kubo formula}
\end{equation}

\subsection{Matsubara Formalism}
\label{sec:MatsubaraFormalism}

In this subsection, we present an alternative derivation of the previous result by implementing the Matsubara formalism, which is based on performing analytical continuation from the Matsubara frequency axis to the real frequency axis. This technique is applicable for systems satisfying the fluctuation dissipation theorem. Within the Matsubara formalism, the correlation function can be expressed as~\cite{AGD, MahanBook}:
\begin{equation}
\chi(q) = \sum_k \operatorname{tr}[G(k) A G(k+q) B].
\label{eq:Matsubara}
\end{equation}
Here, $k \equiv (\omega_n, \bm{k})$ is the internal four-momentum, $q \equiv (\Omega_m, \bm{q})$ is the external four-momentum, and the summation over $k$ denotes both the momentum integration over $\bm{k}$ and the frequency summation over fermionic Matsubara frequencies $\omega_n \equiv (2n+1)\pi T$: $\sum_{k}\equiv T\sum_{n}\sum_{\bm{k}}$. The bosonic Matsubara frequencies are $\Omega_m=2m\pi T$. Taking $A$ and $B$ to be non-dynamical with $A=A^{\dagger}$, $B=B^{\dagger}$, $\Omega \geqslant 0$, the Matsubara summation can be evaluated using contour integration.

In the case of noninteracting systems, the fermionic Matsubara Green's function for a noninteracting system described by PHQM is given below Eq.~\eqref{MatsubaraObserv} of the main text: $G(i \omega_{n}) = (i \omega_{n} - H)^{-1}$. Using the resolution of unity in terms of right and left eigenvectors, and denoting by $n_{F}(\xi)$ the Fermi-Dirac distribution function, Eq.~\eqref{eq:Matsubara} can be evaluated in terms of the pseudo-Hermitian Matsubara Green's function:
\begin{align}
\chi_{A B}(q) &= \text{Tr}[G(k) A G(k+q) B] \nonumber\\
& = \sum_{k}\sum_{\alpha , \beta} \langle L_\alpha| A|R_\beta \rangle\langle L_\beta |B|R_\alpha \rangle \frac{1}{i \omega_n - \xi_\alpha} \frac{1}{i \omega_n + i \Omega_m - \xi_\beta } \nonumber\\
& = \sum_{\bm{k}}\sum_{\alpha \beta} \langle L_\alpha| A|R_\beta \rangle\langle L_\beta|B|R_\alpha \rangle \frac{n_{F}(\xi_\alpha) - n_{F}(\xi_\beta)}{i \Omega_m +\xi_\alpha - \xi_\beta }. 
\end{align}
After performing analytical continuation, $i \Omega_m \rightarrow \Omega + i0^{+}$, the result is
\begin{equation}
\chi_{A B}(q) = \sum_{\bm{k}}\sum_{\alpha \beta} \langle L_\alpha| A|R_\beta \rangle\langle L_\beta|B|R_\alpha \rangle \frac{n_{F}(\xi_\alpha) - n_{F}(\xi_\beta)}{\Omega +\xi_\alpha - \xi_\beta +i0^+}. 
\end{equation}
By incorporating the implicit $\bm{q}$ dependence, we obtain the PHQM response identical to Eq.~\eqref{PHQM Kubo formula}. In deriving this formula we take there to be an infinitesimal decay rate $i0^+$, such that there is no broadening of the spectral function. However, by extending PHQM into a field-theoretic or Green's function-based formalism, we may readily consider Green's functions with finite lifetimes. In the case of a system with finite broadening, performing the Matsubara frequency summation via contour integration gives
\begin{equation}
\begin{aligned}
\chi(\Omega, \bm{q}) &= -\sum_{\bm{k}} \int_{-\infty}^{\infty}  n_{F}(\omega) \operatorname{tr}\biggl[G_R(\omega) A G_R(\omega+\Omega) B-G_A(\omega) A G_R(\omega+\Omega) B \\
& \hspace{3cm} + G_A(\omega-\Omega) A G_R(\omega) B-G_A(\omega-\Omega) A G_A(\omega) B\biggr] \frac{d \omega}{2 \pi i} \\
& =-\sum_{\bm{k}} \int_{-\infty}^{\infty}  n_{F}(\omega) \operatorname{tr}\left[\left(G_R(\omega)-G_A(\omega)\right) A G_{R}(\omega+\Omega) B+G_A(\omega-\Omega) A\left(G_R(\omega)-G_A(\omega)\right) B\right] \frac{d \omega}{2 \pi i}.
\label{eq:MatsubaraExplicit}
\end{aligned}
\end{equation}
 For PHQM with finite broadening, or more generally, while using PHQM in conjunction with dissipative interactions, we evaluate Eq.~\eqref{eq:MatsubaraExplicit}) using the retarded matrix Green's function and the pseudo-metric operator:
\begin{align}
\chi(q) &= -\sum_{\bm{k}}\int^{\infty}_{-\infty} n_{F}(\omega) \text{tr} \biggl\{\left[G_R(\omega) - \eta^{-1}(G_R(\omega))^{\dagger} \eta\right]A G_R (\omega + \Omega)B \nonumber\\
&\quad + \eta^{-1}(G_R(\omega - \Omega))^{\dagger}  \eta A \left[G_R(\omega) - \eta^{-1}(G_R(\omega))^{\dagger} \eta\right] B \biggr\} \frac{d\omega}{2 \pi i}.
\end{align}
\end{widetext}

\section{Distribution Functions} 
\label{DistributionFunctionsSection}

In thermal equilibrium, where the FDT holds, the distribution function can be computed as
\begin{align}
\langle c^{\dagger}_i c_j \rangle &= \langle j| \left(T\sum_{\omega_n} G(i\omega_n)e^{i\omega_n 0^+}\right)|i\rangle \nonumber\\
&=i \langle j|\left(\int_{-\infty}^{\infty}  \left(G_R (\omega) - G_{A}(\omega)\right) n_{F}(\omega)\frac{d \omega}{2 \pi } \right)|i \rangle,
\label{density1}
\end{align}
where $i,j$ are the system degrees of freedom (such as spin and momentum), and $G(k)$, $G_R(\omega)$, and $G_A(\omega)$ are the Matsubara, retarded, and advanced matrix Green's functions respectively. Single-particle observables may be computed from the distribution function as
\begin{equation} 
\langle \mathcal{O} \rangle = \sum_{i j} O_{i j} \langle c^{\dagger}_i c_j \rangle .
\label{observable}
\end{equation}
In the zero-temperature limit, the integral for the distribution function simplifies to 
\begin{equation} 
\langle c^{\dagger}_i c_j\rangle =i \langle j|\left(\int_{-\infty}^{0}  \left(G_R (\omega) - G_{A}(\omega)\right) \frac{d \omega}{2 \pi }\right) |i \rangle.
\label{density2}
\end{equation}
The Green's functions have the forms~\cite{Groenendijk2021}: $G_R(\omega) = \sum_{\alpha} \frac{\hat{\Pi}^R_{\alpha}}{\omega - \xi_{\alpha}}$, $G_A(\omega) = \sum_{\alpha} \frac{\hat{\Pi}^A_{\alpha}}{\omega - \xi^*_{\alpha}}$ , where $\hat{\Pi}^{R/A}_{\alpha}$ is the respective matrix projector and $\xi_{\alpha} \in \mathbb{C}$. The retarded and advanced Green's functions have poles in the lower and upper half planes, respectively, and thus we can evaluate the integral above to obtain 
\begin{equation} \label{density3}
    \langle c^{\dagger}_i c_j\rangle =\frac{i}{2 \pi }  \langle j|\sum_{\alpha} \left(\hat{\Pi}^R_{\alpha} \text{log}(\xi_{\alpha})-\hat{\Pi}^A_{\alpha}\text{log}(\xi^*_{\alpha})\right)|i \rangle,
\end{equation}
where $\sum_{\alpha} \hat{\Pi}^{R/A}_{\alpha} = \mathds{1}$ was used~\cite{Shen2024}. While this calculation has been presented in Ref.~\cite{Shen2024}, here we extend this result to the Green's functions of PHQM. 

For a standard interacting system $\hat{\Pi}^{R}_{\alpha} = |R_{\alpha}\rangle \langle L_{\alpha}|$ and $\hat{\Pi}^{A}_{\alpha} = |L_{\alpha}\rangle \langle R_{\alpha}|$, whereas the retarded and advanced Green's functions of PHQM both have the projector $\hat{\Pi}^{R/A}_{\alpha} = |R_{\alpha}\rangle \langle L_{\alpha}|$. Sharing a common projector, the log$|\xi_{\alpha}|$ terms in $\text{log}(\xi_{\alpha}) = \text{log}|\xi_{\alpha}| + i \text{arg}(\xi_\alpha) $ cancel, and the remaining weights for the projectors depend on arg$(\xi_\alpha)$. By combining Eqs.~\eqref{observable} and \eqref{density3}, we obtain the first two lines of Eq.~\eqref{variousEVs} of the main text -- the zero-temperature expressions for arbitrary (non-dynamical) observables for NH systems in thermal equilibrium. 

Since we are interested in comparing the various approaches to studying NH physics, we also consider postselected dynamics. For postselected dynamics, we consider the NH thermal state (NHTS) motivated and studied in Refs.~\citep{Kou2022, Cao2023, Meden2023}, which applies in the case of real eigenvalues:
\begin{equation}
\label{NHTS}
\rho = \sum_{n}e^{-\beta E_n}|\psi^R_n\rangle \langle \psi^R_n| = e^{-\beta H}\eta^{-1}, 
\end{equation}
Here, the density matrix $\rho$ is expressed as a summation over the projectors formed from the right multi-particle states $|\psi^R_n\rangle$, and the Boltzmann factors in terms of the multi-particle energies $E_n$. By making use of the pseudo-metric operator $\eta$ (in this instance, defined in terms of the right multi-particle states), we may express the NHTS succinctly in terms of the exponential of the NHH and the inverse pseudo-metric operator. The density matrix for the  NHTS is Hermitian and stationary under the generalized von Neumann equation \eqref{generalizedVN}, which describe postselected dynamics. This may be contrasted with the commonly used finite-temperature pseudo-Hermitian density matrix $\rho = e^{-\beta H}$, which is not Hermitian or stationary with respect to Eq.~\eqref{generalizedVN}, rather it is Hermitian with respect to an inner product modified by $\eta$ and stationary for the usual von-Neumann equation.

As described in Ref.~\cite{Kou2022}, in the case of (non-degenerate), complex eigenvalues, where states with positive imaginary eigenvalues are exponentially preferred, the NHTS is set by the state $|\psi^R_0\rangle$ with maximum imaginary part. In which case, $\rho = |\psi^R_0\rangle\langle\psi^R_0|$. In the case of the zero-temperature limit of Eq.~\eqref{NHTS} converging to one of the single-particle right states, $|R_0\rangle\langle R_0|$, then we can denote the ground-state defined by having the minimum real part in the case of real eigenvalues, or maximum imaginary part in the case of complex eigenvalues (assuming no degeneracy), as $|R_0\rangle$. In summary, the equilibrium expectation values are as given in Eq.~\eqref{variousEVs}.

\section{Electrical Conductivity}
\setcounter{subsection}{0}
\subsection{Kubo Formula}

In terms of the response function $\chi_{ij}(\Omega,\bm{q})$, the Kubo formula for optical conductivity is defined as~\cite{Abrikosov1963, Antezza2025}:
\begin{equation}
\sigma_{ij}(\Omega,\bm{q}) = \frac{i}{\Omega + i0^+}\left[\chi_{ij}(\Omega,\bm{q})-\lim_{\Omega\rightarrow0}\chi_{ij}(\Omega,\bm{q})\right].
\label{GeneralKuboFormula}
\end{equation}
When the response function is regularized by nonzero inverse lifetimes, the order of the limits $\bm{q}\rightarrow0$ and $\Omega\rightarrow0$ commute. The local optical conductivity is then given by
\begin{align}
\sigma_{ij}(\Omega) &\equiv \lim_{\bm{q}\rightarrow0}\sigma_{ij}(\Omega,\bm{q})\\\nonumber &= \frac{i}{\Omega + i0^+}\left[\lim_{\bm{q}\rightarrow0}\chi_{ij}(\Omega, \bm{q})-\lim_{\bm{q}\rightarrow0}\lim_{\Omega\rightarrow0}\chi_{ij}(\Omega, \bm{q})\right],
\label{LocalKubo}
\end{align}
Let $\chi_{ij}(\Omega) \equiv \underset{\bm{q}\rightarrow 0}{\lim}\chi_{ij}(\Omega,\bm{q})$. Applying the Sokhotski–Plemelj formula gives
\begin{equation}
\begin{aligned}
\sigma_{ij}(\Omega) &= \mathcal{P}\frac{i}{\Omega}\left[\chi_{ij}(\Omega)-\lim_{\bm{q}\rightarrow0}\lim_{\Omega\rightarrow0}\chi_{ij}(\Omega, \bm{q})\right]\\  &+ 
\pi \delta(\Omega)\left[\chi_{ij}(\Omega) -\lim_{\bm{q}\rightarrow0}\lim_{\Omega\rightarrow0}\chi_{ij}(\Omega, \bm{q})\right],
\end{aligned}
\label{RegularizedConductivity}
\end{equation}
where $\mathcal{P}$ denotes the Cauchy principal value. The second term is zero; thus, there is no Drude weight in the system~\cite{DupuisBook}. Hence, only the regular part contributes to the conductivity. 

Integrating the real part of the conductivity over frequency and applying the Kramers-Kronig relation gives the conductivity sum rule~\cite{MahanBook, DupuisBook}:
\begin{equation}
\mathcal{P}\int_{-\infty}^{\infty}\sigma^{\prime}(\Omega)d\Omega = -\pi\lim_{\Omega\rightarrow0}\lim_{\bm{q}\rightarrow0} \chi^{\prime}(\Omega,\bm{q}).
\label{eq:OSR}
\end{equation}
The prime notation corresponds to the real part of a complex function, and we consider only the $x$ component of the conductivity tensor. The systems we consider have dissipation present (reflected in the nonzero lifetime in the retarded Green's function). In this case, the order of limits commute~\cite{DupuisBook} and thus
\begin{equation}
\lim_{\Omega\rightarrow0} \lim_{\bm{q}\rightarrow0} \chi^{\prime}(\Omega,\bm{q}) = \lim_{\bm{q}\rightarrow0} \lim_{\Omega\rightarrow0} \chi^{\prime}(\Omega,\bm{q}).
\end{equation}
For brevity, we shall simply write $\chi^{\prime}(0)$ from now on, and it is understood that the limits commute for the model(s) under consideration. A more general discussion of sum rules can be found in Ref.~\cite{DupuisBook}. 

\begin{widetext}
\subsection{DC conductivity computed using the Matsubara formalism}

\label{sec:Conductivity}
For a system with nonzero inverse quasiparticle lifetime, the local conductivity is given by substituting Eq.~\eqref{eq:MatsubaraExplicit} into Eq.~\eqref{RegularizedConductivity}. The (temperature-dependent) DC conductivity is determined to be
\begin{equation}
\begin{aligned}
\sigma_{\DC}(T)&=-\lim _{\Omega \rightarrow 0} \frac{1}{\Omega} \sum_{\bm{k}} \int_{-\infty}^{\infty} \{\left( n_F ( \omega + \Omega ) - n_F ( \omega ) \right) \operatorname{tr}\left[G_{A}(\omega) A G_{R}(\omega+\Omega) B\right] \\
&\quad +n_{F}(\omega) \operatorname{tr}\left[G_{R}(\omega) A\left(G_{R}(\omega+\Omega)-G_{R}(\omega)\right) B-\left(G_{A}(\omega-\Omega)-G_{A}(\omega)\right) A G_{A}(\omega) B\right]\}\frac{d \omega}{2 \pi} \\ 
&=-\sum_{\bm{k}} \int_{-\infty}^{\infty} \left\{\partial_\omega n_{F}(\omega) \operatorname{tr}\left[G_{A}(\omega) A G_{R}(\omega) B\right]+n_{F}(\omega) \operatorname{tr}\left[G_{R}(\omega) A\left(\partial_\omega G_{R}(\omega)\right) B+\left(\partial_\omega G_{A}(\omega)\right) A G_{A}(\omega) B\right]\right\} \frac{d \omega}{2 \pi}. 
\end{aligned}
\end{equation}
In the zero-temperature limit, we obtain
\begin{align}
\sigma_{\DC} &=\sum_{\bm{k}} \int_{-\infty}^{\infty} \left\{\delta(\omega) \operatorname{tr}\left[G_{A}(\omega) A G_{R}(\omega) B\right]-\theta(-\omega) \;\text{tr}\left[G_{R}(\omega) A\left(\partial_\omega G_{R}(\omega)\right) B+\left(\partial_\omega G_{A}(\omega)\right) A G_{A}(\omega) B\right]\right\}\frac{d \omega}{2 \pi} \nonumber \\
\label{DC_cond_omega}
 &=  \sum_{\bm{k}} \left\{\operatorname{tr}\left[G_{A}(0) A G_{R}(0) B\right]- \int_{-\infty}^0 \operatorname{tr}\left[G_{R}(\omega) A\left(\partial_\omega G_{R}(\omega)\right) B+\left(\partial_\omega G_{A}(\omega)\right) A G_{A}(\omega) B\right]\frac{d \omega}{2 \pi}\right\}.
\end{align}
When the retarded or advanced Green's functions are of the form $G(\omega)=(\omega-\mathcal{O})^{-1}$, for a frequency-independent operator $\mathcal{O}$, we differentiate the identity $G(\omega)G^{-1}(\omega)=\mathds{1}$ with respect to $\omega$ to obtain $\partial_\omega G = -(G(\omega))^2$. For a longitudinal response $A=B=J\equiv -e\partial_{k_i} G^{-1}$. Note that $\partial_{k_i}\left(G G^{-1}\right)=\left(\partial_{k_i} G\right) G^{-1}+G\left(\partial_{k_i} G^{-1}\right)=0$, where in the absence of an explicit argument, the Green's functions are taken to be momentum- and frequency-dependent. Thus,
\begin{equation}
\operatorname{tr}[G\left(\partial_{k_i} G^{-1}\right)\left(\partial_\omega G\right)\left(\partial_{k_i} G^{-1}\right)] =- \operatorname{tr}[G\left(\partial_{k_i} G^{-1}\right) G^2\left(\partial_{k_i} G^{-1}\right)]  
=\frac{1}{2}\operatorname{tr}[\left(\partial_{k_i} G^{-1}\right) \left(\partial_{k_i} G^2\right)].
\end{equation}
Using the above identity, we can simplify Eq.~\eqref{DC_cond_omega} to obtain the longitudinal DC conductivity:
\begin{equation}
\sigma_{\DC}=\frac{1}{2 \pi} \sum_{\bm{k}} \left\{\operatorname{tr}\left[G_{A}(0) J G_{R}(0) J\right] - \frac{e^2}{2}\int_{-\infty}^{0} \operatorname{tr}\left[\left(\partial_{k_{i}} G_R^{-1}\right)\left(\partial_{k_{i}} G_R^{2}\right)+(\partial_{k_{i}}G^{-1}_{A})(\partial_{k_{i}}G_{A}^{2})\right] d\omega\right\}.
\label{DC_general}
\end{equation}
In standard systems, where $G_R=G_A^{\dagger}$, the conductivity simplifies to:
\begin{equation}
\begin{aligned}
\sigma_{\DC} &= \frac{1}{2 \pi} \sum_{\bm{k}}\left\{\operatorname{tr}\left[G_{A}(0) J G_{R}(0) J\right] - e^2\int_{-\infty}^{0}  \operatorname{Re} \operatorname{tr}\left[\left(\partial_{k_{i}} G_R^{-1}\right) \left(\partial_{k_{i}} G_R^2\right)\right]d\omega\right\}\\
&=\frac{1}{2 \pi} \sum_{\bm{k}}\left\{\operatorname{tr}\left[G_{A}(0) J G_{R}(0) J\right] + e^2\int_{-\infty}^{0}  \operatorname{Re} \operatorname{tr}\left[\left(\partial_{k_{i}}^2 G_R^{-1}\right) G_R^2\right] d\omega \right\}.
\end{aligned}
\label{eq:DC_standard}
\end{equation}
In the second step, we used integration by parts and dropped a boundary term that does not contribute. For systems where $\partial_{k_i}J\sim\partial_{k_i}^2G^{-1}_{R}=0$, in the continuum limit, the second term vanishes and the first term is the sole contribution to the DC conductivity. Momentum-independent current operators are common in linear or Dirac-like systems.

\subsection{Applying the Matsubara Formalism to a Standard System}
In this subsection, we compute the DC conductivity for the Tachyon model, where the non-Hermitian terms in the quasiparticle Hamiltonian are identified with the anti-Hermitian part of the self-energy. The Matsubara Green’s function is given by 
\begin{align}
    G(i\omega_n)=\left(i\omega_n - (v_Fk\sigma_x + \Delta \sigma_y -\mu \mathds{1})+i\text{sgn}(\omega_n)(m\sigma_z+\gamma\mathds{1})\right)^{-1}; \quad   \gamma>|m|,\;\gamma>0.
\end{align}
In the particle-hole symmetric case, where $\mu=0$, the retarded Green's function is given by
\begin{align}
G_R(\omega) = 
\begin{pmatrix}
\omega + i (\gamma+m)  & -v_F k + i \Delta \\
-v_F k - i \Delta & \omega +i (\gamma-m) 
\end{pmatrix}^{-1}.
\label{RetardedGF}
\end{align}
The advanced Green's function is $G_A=G_R^{\dagger}$. The spectral properties of this model are shown in Fig.~\ref{fig:SpectralFn} in terms of the spectral function $A(\omega,k)\equiv i\operatorname{tr}\left[G_{R}(\omega,k)-G_{A}(\omega,k)\right]$. 

\begin{figure}[h]
\centering\includegraphics[width=1\columnwidth]{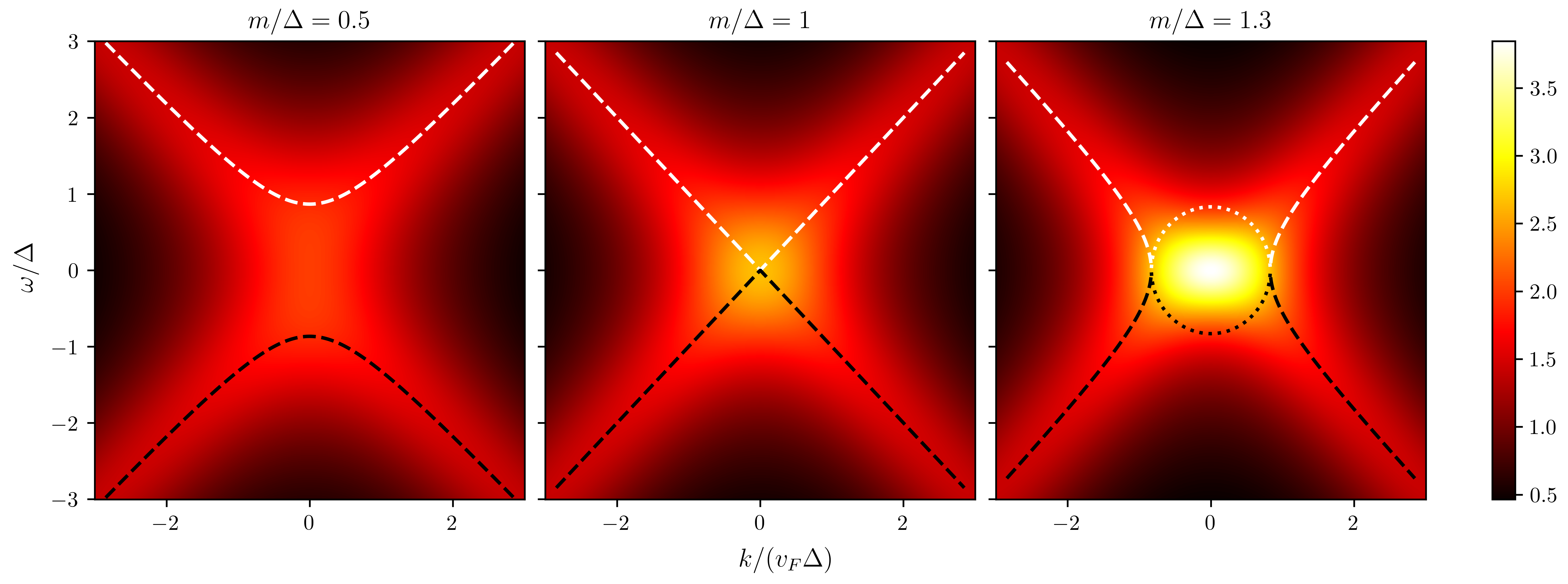}
\caption{Fermionic spectral function in units of $\Delta^{-1}$, for $\mu=0$ and $\gamma=1.5\Delta$, overlaid with the energy eigenvalues. The black and white dashed lines denote the valence and conduction bands respectively, determined from the real part of the spectrum. The dotted lines in the third figure ($m=1.3\Delta$) denote the imaginary part of the spectrum.}
\label{fig:SpectralFn}
\end{figure}

The current operator is given by $J=-e\partial G^{-1}/\partial k = ev_F \sigma_x$.
Substituting the retarded and advanced matrix Green's functions along with the current operator $J=e v_F \sigma_x$ into Eq.~\eqref{eq:DC_standard}, we obtain the following result:
\begin{align}
\sigma_{\DC} &= \frac{e^2 v_F^2}{\pi}\int_{-\infty}^{\infty} \frac{\gamma^2+v_F^2k^2 -\Delta^2-m^2}{(\gamma^2+v_F^2k^2 +\Delta^2-m^2)^2}\frac{dk}{2\pi} \nonumber\\
&= \frac{e^2v_F}{2\pi}\frac{\gamma^2-m^2}{\left(\gamma^2+\Delta^2-m^2\right)^{3 / 2}}.
\end{align}
While we consider arbitrary values of $m/\Delta$ here, we take $\gamma > |m|$ to ensure dynamical stability. Taking the zero-temperature and zero-frequency limit of Eq.~\eqref{eq:MatsubaraExplicit}:
\begin{align}
 -\pi \chi^{\prime}(0)|_{T=0,\mu=0}
& = \sum_{\bm{k}} \int_{-\infty}^{0} \text{Im}\text{Tr}\left[G_R(\omega)JG_R(\omega)J\right] d \omega \nonumber \\ 
&= e^2v_F^2\int_{-\infty}^{\infty}\left\{ \int_{-\infty}^{0}  \text{Im}\left[ \frac{2((\omega+i\gamma)^2+v_F^2k^2-\Delta^2+m^2)}{(-(\omega+i\gamma)^2+v_F^2k^2+\Delta^2-m^2)^2}\right] d \omega\right\} \frac{dk}{2\pi} \nonumber \\
&= e^2 v_F.
\label{StdMatsubara3}
\end{align}

\subsection{Applying the Matsubara formalism to PHQM}
\label{PHQMConductivity1}

In this subsection, we compute the DC conductivity for the Tachyon model using PHQM. 
The Hamiltonian is $H=v_F k \sigma_x+\Delta \sigma_y-i m \sigma_z-\mu \mathds{1}$. 
The Matsubara Green's function for a system described within PHQM is given by
\begin{equation}
G(i\omega_n)=(i\omega_n-H+i \operatorname{sgn}(\omega_n) \gamma)^{-1}; \quad  \gamma>0, \quad H \neq H^{\dagger}. 
\end{equation}
This Green's function is different from the one within an interacting theory in that the anti-Hermitian part of $H$ is not accompanied by $\text{sgn}(\omega_n)$. The term $-i \gamma \mathds{1}$ (responsible for finite quasiparticle lifetime) is not pseudo-Hermitian and is accompanied by the signum function. Unlike the non-Hermiticity in $H$, $\gamma$ describes a decay rate arising from a dissipative interaction. As in the standard treatment of an interacting system, where the quasi-particle Hamiltonian need not be Hermitian, the quasi-particle Hamiltonian of a system described by PHQM need not be pseudo-Hermitian. It should be emphasized that although we are considering the same quasiparticle Hamiltonian (described by a NHH $H$ and a uniform lifetime), the underlying dynamics are distinct from the previous calculation.

In the previous calculation, we assumed that the anti-Hermitian part of the quasiparticle Hamiltonian arose from a standard dissipative interaction, and thus it was accompanied by a signum function. Here, we will take the NH quasiparticle Hamiltonian to be described by PHQM, where it is equivalently described by an isospectrally related Hermitian Hamiltonian $h$. The pseudo-Hermitian portion of the quasiparticle Hamiltonian prescribes a Hermitian positive-definite quantum metric $\eta$ which endows the Hilbert space for the NHH. The square root of the pseudo metric operator defines the similarity transformation that relates the NHH to its isospectral counterpart $h$. As a result, the projectors for the advanced Green's function are the same as those for the retarded. Consequently, the anti-Hermitian and pseudo-Hermitian portion of $H$ is not accompanied by a signum function.

The retarded Green's function is given in Eq.~\eqref{RetardedGF}. However, the advanced Green's function is not the adjoint of the retarded Green's function: $G_A\neq G_R^{\dagger}$. Instead, the correct relation is $G_{A} = \eta^{-1} (G_{R})^{\dagger} \eta$. Using the intertwining relation, $H^{\dagger}=\eta H \eta^{-1}$, the advanced Green's function is given by $G_A(\omega)=(\omega-H-i\gamma)^{-1}$, and in matrix form it is
\begin{equation}
G_A(\omega) = 
\begin{pmatrix}
\omega + i (-\gamma+m)  & -v_F k + i \Delta \\
-v_F k - i \Delta & \omega -i (\gamma+m) 
\end{pmatrix}^{-1}.
\end{equation}
Once again, we set $\mu=0$. Substituting the retarded and advanced matrix Green's functions and the current operator $J=e v_F \sigma_x$ into Eq.~\eqref{DC_general}, and then taking the continuum limit, we obtain the following result:
\begin{align}
\sigma_{\DC} &= \frac{e^2 v_F^2}{\pi}\int_{-\infty}^{\infty} \frac{\gamma^2+v_F^2k^2 -\Delta^2+m^2}{(\gamma^2+v_F^2k^2 +\Delta^2-m^2)^2}\frac{dk}{2\pi}  \nonumber\\
&= \frac{e^2v_F}{2\pi}\frac{\gamma^2}{\left(\gamma^2+\Delta^2-m^2\right)^{3 / 2}}.
\end{align}
In this approach, we restrict ourselves to $\Delta^2>m^2$ to ensure real eigenvalues and a positive-definite pseudo-metric operator. Since the non-Hermitian part of $H$ is not interpreted as a self-energy, positive definiteness ($\gamma>0$) is the only restriction for the decay rate. 

Taking the zero-temperature and zero-frequency limit of Eq.~\eqref{eq:MatsubaraExplicit}, we evaluate the right-hand side of the optical sum rule in Eq.~\eqref{eq:OSR}. The result is given by
\begin{align}
&-\left.\pi \chi^{\prime}(0)\right|_{T=0, \mu=0} \nonumber\\
&=\frac{1}{2}\sum_{\bm{k}}\int_{-\infty}^{0}\text{Im}\left[G_R(\omega)JG_R(\omega)J-G_A(\omega)JG_A(\omega)J\right]\ d\omega \nonumber\\ 
&=4e^2v_F \sum_{\bm{k}}\int_{-\infty}^{0} \gamma\omega\biggl[-\gamma^4 + 2\gamma^2 (-\omega^2 + v_F^2 k^2 - \Delta^2 + m^2) - (\omega^2 - v_F^2 k^2 - \Delta^2 + m^2) (\omega^2 + 3 v_F^2 k^2 - \Delta^2 + m^2)\biggr] \nonumber\\
&\quad\quad \times \biggl[ \gamma^4 + 2 \gamma^2 (\omega^2 + v_F^2 k^2 + \Delta^2 - m^2) + (\omega^2 - v_F^2 k^2 - \Delta^2 + m^2)^2\biggr]^{-2}\ d\omega \nonumber\\
&=e^2v_F.
\end{align}

\subsection{Applying the Matsubara Formalism to the Isospectral System}
\label{App:IsospecMatsubara}

Conductivity computed using PHQM with the current $J$ merely leverages unitary time-evolution (via a positive-definite metric). To achieve equivalence with the Hermitian system $h$, the isospectral transformation must be applied to observables~\cite{Mostafazadeh2007}. When the isospectral transformation is applied to the current derived from $h$ to obtain $\widetilde{j}$, the response evaluated for $H$ is identical to that of $h$. For this particular model, the velocity operator for the transformed current 
$\widetilde{j}\equiv-e\widetilde{\bm{v}}(\bm{k})$, evaluates to $\widetilde{\bm{v}}(\bm{k})= v_F (\bm{a}\cdot \bm{\sigma})$, in terms of the Bloch vector $\bm{a}$, given by
\begin{align}
\bm{a} \doteq \left(v_F^2 k^2E^{-2}+\Delta^2EE_0^{-3}, v_F k \Delta (E^{-2}-E E_0^{-3}),-i v_F k m E^{-2}\right),
\end{align}
where $E_0 \equiv E|_{m=0}$, and $\bm{a}|_{m=0} \doteq (1,0,0)$. The current $\widetilde{j}$, is computed from the isospectral Hamiltonian and $\eta^{1/2}$, and for nonzero imaginary mass $m$, develops $\sigma_{y/z}$ components of the Bloch vector $\bm{a}$.

The current-current response for the isospectral system may be computed equivalently in one of two ways. We may compute the response in terms of NHH $H$ (as done in Subsection~\ref{PHQMConductivity1}) with the transformed current $\widetilde{j}$ from the main text, or we may compute conductivity in terms of the Hermitian isopsectral Hamiltonian $h$ and its respective current $j$. Here we show the details for the latter. The Matsubara Green's function is defined as
\begin{equation}
g(i\omega_n) = (i \omega_n - h + i \text{sgn}(\omega_n)\gamma)^{-1},
\end{equation}
From the Green's function, the current operator is then given by $j = -e\partial_{k}g^{-1}(i\omega_{n})$. Let $\Omega_{m} \geq 0$. The real-frequency current-current response is then
\begin{equation}
\begin{aligned}
\chi(q) &= \sum_k \operatorname{tr}[g(k) j g(k+q) j]\\
& =-\sum_{\bm{k}} \int_{-\infty}^{\infty} \frac{d \omega}{2 \pi i} n_{F}(\omega) \operatorname{tr} \left[\left( g_{R}(\omega) - g_{A}(\omega)\right)j g_R(\omega+\Omega) j + g_A(\omega-\Omega) j\left(g_R(\omega)-g_A(\omega)\right) j\right],
\end{aligned}
\end{equation}
where $g_R(\omega) = (\omega - h + i \gamma)^{-1}$ and $g_A(\omega) = (\omega - h - i \gamma)^{-1} = \left(g_{R}(\omega)\right)^{\dagger}$ are the respective retarded and advanced Green's functions for the isospectral Hamiltonian. 
The DC conductivity is computed with Eq.~\eqref{RegularizedConductivity}. Both terms in Eq.~\eqref{eq:DC_standard} contribute, since $\partial_{k}j\neq0$. Taking the $\Omega\rightarrow 0$ limit, then integrating over the internal frequency $\omega$, we obtain
\begin{equation}
\sigma^{\prime} = e^2v_F^2\sum_{\bm{k}} \frac{2\gamma^2}{\pi}\frac{(v_F^2k^2+\Delta^2)^3-2\Delta^2m^2(v_F^2k^2+\Delta^2)+\Delta^2m^4}{(v_F^2k^2+\Delta^2)^2(v_F^2k^2+\Delta^2-m^2)(v_F^2k^2+\Delta^2-m^2+\gamma^2)^2}.
\end{equation}
Performing the momentum integration gives the following result:
\begin{align}
\sigma^{\prime} &= \frac{e^2v_F}{2\pi\gamma^2 \left(\gamma^2-m^2\right)^3}\biggl[\frac{\gamma^4 \left(\gamma^2 \left(2 \Delta^2-m^2\right)+2 \Delta^2 m^2+m^4\right)}{|\Delta|}-2 \sqrt{\Delta^2-m^2}\left(\gamma^2-m^2\right)^3 \nonumber\\ 
&\quad +  \left(\gamma^2+\Delta^2-m^2\right)^{-3/2}\biggl(\gamma^{10} 
 + m^8 \left(9 \gamma^2+4 \Delta^2\right)-m^6 \left(16 \gamma^4+15 \gamma^2 \Delta^2+2 \Delta^4\right) \nonumber\\
 & \quad +\gamma^2 m^4 \left(14 \gamma^4+23 \gamma^2 \Delta^2+6 \Delta^4\right) -2 \gamma^4 m^2 \left(3 \gamma^4+6 \gamma^2 \Delta^2+4 \Delta^4\right)-2 m^{10}\biggr)\biggr].
\end{align}
The clean and dirty expansions (in $\gamma$), are evaluated and reported in the main text. We also evaluate the zero-temperature OSR:
\begin{equation}
\begin{aligned}
 -\left.\pi \chi^{\prime}(0)\right|_{T=0,\mu=0}
& = \sum_{\bm{k}} \int_{-\infty}^{0} \text{Im}\operatorname{tr}\left[g_R(\omega)jg_R(\omega)j \right] d \omega \\ 
&=\frac{e^2 v_F}{2}\Biggl[1 + \gamma \left(\frac{2}{\sqrt{\Delta^2-m^2}+\sqrt{\gamma^2+\Delta^2-m^2}}-\frac{1}{\Delta + \sqrt{\gamma^2+\Delta^2-m^2}}\right) \\ & \quad + \left(\frac{\Delta}{m}-\frac{m}{\Delta}\right)\left(\mathrm{artanh}\left(\frac{m}{\gamma}\right)+\mathrm{artanh}\left(\frac{m}{\Delta}\right)-\mathrm{artanh}\left(\frac{m \sqrt{\gamma^2+\Delta^2-m^2}}{\gamma \Delta}\right)\right)
 \Biggr].
\end{aligned}
\end{equation}
In the above calculation, we assumed that $\Delta>0$, $\Delta^2 > m^2 > 0$, and $\gamma > 0$. Note that, there is no restriction on the size of $\gamma/|m|$; that is, one can consider $\gamma>|m|$ in analogy with the requirement of positive definiteness of $\Sigma_{R}^{\prime\prime}$ in the interacting case, and one can also consider the case where $|m|>\gamma$. 

Define the dimensionless variables $\bar{m}\equiv m/\Delta$ and $\bar{\gamma} \equiv \gamma/\Delta$. In the weak non-Hermiticity limit, where $\bar{m}\ll1$ and $\bar{m}\ll\bar{\gamma}$, the optical sum can be expanded in powers of $\bar{m}$ to give:
\begin{equation}
 -\pi \chi^{\prime}(0)|_{T=0,\mu=0}
 = e^2v_F \left\{1 + \left[2 - 2 \sqrt{\bar{\gamma}^2+1}+\bar{\gamma}^2\left(\sqrt{\bar{\gamma}^2+1}-\bar{\gamma}\right)\right]\frac{1}{3\bar{\gamma}^3}\bar{m}^2 + \mathcal{O}(\bar{m}^{4})\right\}.
\end{equation}
The limit of strong non-Hermiticity corresponds to $|\bar{m}| \rightarrow 1^{-}$. In this limit, the optical sum is
\begin{equation}
-\left.\pi \lim_{|\bar{m}|\rightarrow1^{-}}\chi^{\prime}(0)\right|_{T=0,\mu=0} = e^2v_F \left[1 + \frac{1}{2(1+\bar{\gamma})}\right].
\end{equation}
We can then subsequently take the limit that $\bar{\gamma}\rightarrow0$. Alternatively, one can consider the opposite order of limits and first take the limit $\bar{\gamma}\rightarrow0$ and then consider $|\bar{m}|\approx1^{-}$. 
In the ``clean'' case, as $\bar{\gamma}\rightarrow0$, the sum rule evaluates to:
\begin{equation}
-\left.\pi \lim_{\bar{\gamma}\rightarrow0}\chi^{\prime}(0)\right|_{T=0,\mu=0}
 = \frac{1}{2}e^2v_F \left[1 + \left(\bar{m}^{-1}-\bar{m}\right)\mathrm{artanh}\left(\bar{m}\right)\right].
\end{equation}
Taking $|\bar{m}|\rightarrow 1^{-}$ followed by $\bar{\gamma}\rightarrow 0$ in the optical sum rule corresponds to considering a strong non-Hermiticity parameter with a small uniform inverse lifetime, while the reverse order $\bar{\gamma}\rightarrow 0$ followed by $|\bar{m}|\rightarrow 1^{-}$ corresponds to a clean system with strong non-Hermiticity. Notice that the order of limits $|\bar{m}|\rightarrow 1$ and $\bar{\gamma}\rightarrow 0$ does not commute:
\begin{align}
-\left.\pi \lim_{\bar{\gamma}\rightarrow0}\lim_{|\bar{m}|\rightarrow 1^{-}}\chi^{\prime}(0)\right|_{T=0,\mu=0}
 &= \frac{3}{2}e^2v_F, \\
    -\left.\pi \lim_{|\bar{m}|\rightarrow 1^{-}}\lim_{\bar{\gamma}\rightarrow0}\chi^{\prime}(0)\right|_{T=0,\mu=0}
 &= \frac{1}{2}e^2v_F.
\end{align}
Thus, the optical sum for a system with strong non-Hermiticity is enhanced in the clean limit, whereas it is diminished in a clean system in the strong non-Hermiticity limit. The optical conductivities displayed in the main text reflect the increased weight in the optical sum for $|\bar{m}| \rightarrow 1^{-}$, with $\gamma$ fixed and nonzero.

The imaginary mass $m$ is a tuning parameter for the transition between an insulating phase with a direct band gap at the $\Gamma$ point ($k=0$) and a gapless hyperbolic spectrum with exceptional points. When $m=\Delta$ the system has a linear Dirac-like spectrum with an exceptional point at $k=0$. Reference~\cite{Rivero2023} explores the differences between an exceptional Dirac point and a Dirac point.
\end{widetext}

\section{Additional Derivations}
\setcounter{subsection}{0}
\subsection{Unitary Time Evolution within PHQM}

In the following subsections, we present additional details that are pertinent to the discussion in the main text. As discussed previously, in closed quantum mechanical systems described by a non-Hermitian Hamiltonian, having real spectra is an important property, but ensuring that the dynamics are unitary is another crucial feature~\cite{Meden2023}. If one uses the standard inner product in quantum mechanics, then even if the energy eigenvalues are real, the time evolution is still nonunitary. However, within the formalism of PHQM~\cite{Mostafazadeh2010}, time evolution is manifestly unitary. Here we prove the result in Eq.~\eqref{real-time-obs} of the main text, namely, $\langle O(t)\rangle = \langle e^{iHt}\widetilde{O}e^{-iHt}\rangle_{\eta }$. 

Consider a finite-dimensional Hilbert space $\mathcal{H}$ with a complete orthonormal basis $\{|\phi_n\rangle\}$. Let $|\psi\rangle$ be a normalized state in $\mathcal{H}$ and define $\langle\psi|O|\psi\rangle \equiv \sum_{nm}c_{nm}\langle\phi_n|O|\phi_m\rangle $, where $c_{nm}$ are coefficients that can be deduced from expanding $|\psi\rangle$ and $\langle\psi|$ in terms of $\{|\phi_n\rangle\}$ basis (we omit using the hat notation for operators). By definition, we have
\begin{align}
\langle O(t)\rangle &= \langle e^{iht}Oe^{-iht}\rangle \nonumber\\
&= \sum_{nm}c_{nm}\langle \phi_n|e^{iht}Oe^{-iht} |\phi_m\rangle \nonumber\\
&=\sum_{nm}c_{nm}\langle \phi_n| \eta^{1/2}e^{iHt}\eta^{-1/2}O \eta^{1/2}e^{-iHt}\eta^{-1/2}|\phi_m\rangle \nonumber \\
&= \sum_{nm}c_{nm} \langle \phi_n| \eta^{1/2}e^{iHt}\widetilde{O}e^{-iHt}\eta^{-1/2}|\phi_m\rangle. 
\end{align}
In the last step, we have defined the transformed observable $\widetilde{O}\equiv \eta^{-1/2}O\eta^{1/2}$. Next, define $|\psi_m\rangle \equiv \eta^{-1/2}|\phi_m\rangle$. Then, using the Hermiticity of $\eta^{1/2}$, we have $\langle\psi_n| \equiv \langle \phi_n|\eta^{-1/2}$ $\implies \langle\psi_n| \eta \equiv \langle \phi_n|\eta^{1/2}$. Therefore,
\begin{align}
\langle O(t)\rangle &= \sum_{nm}c_{nm} \langle \phi_n| \eta^{1/2}e^{iHt}\widetilde{O}e^{-iHt}\eta^{-1/2}|\phi_m\rangle \nonumber \\
&= \sum_{nm}c_{nm} \langle \psi_n| \eta e^{iHt}\widetilde{O}e^{-iHt}|\psi_m\rangle \nonumber \\
&\equiv \langle e^{iHt}\widetilde{O}e^{-iHt}\rangle_{\eta }.
\end{align}

Notice that the states $\{|\phi_n\rangle\}$ that are orthonormal under the standard inner product (in $\mathcal{H}$) are mapped onto the states $\{|\psi_n\rangle\}$ that are orthonormal under the inner product endowed with the metric $\eta$ (in $\mathcal{H}_{\eta}$). If we take $\{|\phi_n\rangle\}$ to be the eigenstates of $h$, then we can establish $\{|\psi_n\rangle\}$ as the right eigenstates of $H$ and $\{\eta|\phi_n\rangle\}$ as the left eigenstates. This can be demonstrated as follows. In terms of the $\{|\phi_n\rangle\}$ basis, the hermitian operator $h$ can be expressed as $h = \sum_{n}E_n |\phi_n\rangle\langle\phi_n|.$ Since $H = \eta^{-1/2}h\eta^{1/2}$, then
\begin{equation}
H = \sum_{n}E_n \eta^{-1/2}|\phi_n\rangle\langle\phi_n|\eta^{1/2} = \sum_{n}E_n |\psi_n\rangle\langle\psi_n|\eta.
\end{equation}
The biorthonormal decomposition of a NHH in terms of its left and right eigenstates is $H = \sum_{n}E_{n}|R_n\rangle\langle L_n|$, thus, we identify $|\psi_n\rangle = |R_n\rangle$ and $\eta |\psi_n\rangle = |L_n\rangle$ when $\{|\phi_n\rangle\}$ are the eigenstates of $h$.

\subsection{Green's Functions within PHQM}
Here we provide additional details on the proof that $g = \eta^{1/2}G\eta^{-1/2}$, where $h = \eta^{1/2}H\eta^{-1/2}$, as discussed below Eq.~\eqref{MatsubaraObserv} in the main text. Using the definition of Matsubara Green's function $g$ for the Hermitian Hamiltonian $h$:
\begin{align}
g(i\omega_{n}) &= \left(i\omega_n-h\right)^{-1} \nonumber\\
&= \left( \eta^{1/2}(i\omega_n-H)\eta^{-1/2}\right)^{-1} \nonumber\\
&= \eta^{1/2}\left( i\omega_n-H\right)^{-1}\eta^{-1/2} \nonumber\\
&= \eta^{1/2}G(i\omega_{n})\eta^{-1/2},
\end{align}
where we used $(ABC)^{-1} = C^{-1}B^{-1}A^{-1}$ and the Matsubara Green's function in PHQM defined as $G = (i\omega_n - H)^{-1}$. 

Next, we establish the relationship between the retarded and advanced Green's functions in PHQM. The advanced Green's function is defined by $G_{A}(\omega) = \left(\omega - H - i0^{+}\right)^{-1}$. Thus, $(G_{A}(\omega))^{\dagger} = \left(\omega - H^{\dagger} +i 0^{+}\right)^{-1}$, since the adjoint and inverse operations commute. Using properties of matrix inversion, we obtain the following result:
\begin{align}
G_{R}(\omega) &= \left(\omega - H + i0^{+}\right)^{-1}  \nonumber\\
&= \left(\eta^{-1}(\omega - H^{\dagger} + i0^{+})\eta \right)^{-1} \nonumber\\
&= \eta^{-1}\left(\omega - H^{\dagger} + i0^{+}\right)^{-1}\eta \nonumber\\
&=\eta^{-1}(G_{A}(\omega))^{\dagger}\eta. 
\end{align}
This result can be contrasted with the usual result in quantum field theory: $G_{R}(\omega)=(G_{A}(\omega))^{\dagger}$, which holds for standard interacting systems with the conventional inner product ($\eta=\mathds{1}$). For such systems, the real-time Green's functions are defined as~\cite{MahanBook}:
\begin{align}
G_{R,ij}(t,t^\prime) &= -i \Theta(t-t^\prime)\langle[c_{i}(t),c^{\dagger}_{j}(t^\prime)]_{\pm}\rangle,  \\
G_{A,ij}(t,t^\prime) &= i \Theta(t^\prime-t)\langle[c_{i}(t),c^{\dagger}_{j}(t^\prime)]_{\pm}\rangle.
\end{align}
The elements of $(G_{A})^\dagger$ are
\begin{align}
(G_{A,ji}(t^\prime,t))^{*} &= (i \Theta(t-t^\prime)\langle[c_{j}(t^\prime),c^{\dagger}_{i}(t)]_{\pm}\rangle)^* \nonumber \\
&=-i \Theta(t-t^\prime)\langle[c_{i}(t),c^{\dagger}_{j}(t^\prime)]_{\pm}\rangle \nonumber\\
&=G_{R ij}(t,t^\prime).
\end{align}
Here we have taken the full \emph{interacting} Hamiltonian to be Hermitian and used $\langle\cdot\rangle^* = \langle\cdot^\dagger\rangle$, which holds when density matrix the expectation value is performed over is Hermitian. In which case $G_{R}=(G_A)^{\dagger}$. For a NH system described by PHQM, neither of these conditions hold.

\subsection{Expectation Values within PHQM}
In general, the Hamiltonian corresponds to the generator of time translations. For a Hamiltonian $H$, the time-evolution operator is $U(t)\equiv e^{-iHt}$; in the case where $H$ is NH, $U^\dagger(t)U(t)\neq \mathds{1}$ and thus $U(t)$ is not a unitary linear operator. However, if one endows the Hilbert space with the pseudo-metric operator,  then $U(t)$ becomes a unitary linear operator. 
\begin{enumerate}
\item For all $t\in\mathbb{R}$, $U(t)$ is a bijective linear operator $U(t): \mathcal{H}\rightarrow \mathcal{H}$, where $\mathcal{H}$ is a Hilbert space.
\item $ \langle U(t)\psi , U(t) \phi \rangle_\mathcal{H} = \langle \psi ,\phi \rangle_\mathcal{H}$,  $\forall \; \psi , \phi \in \mathcal{H} .$
\end{enumerate}
The first requirement holds since $U(t)$ is invertible $(U(t))^{-1}= e^{iHt}$, and all invertible functions are bijective. Furthermore, the second requirement holds when we define the inner product as
\begin{equation}
\langle \psi,O \phi \rangle \equiv \langle \psi|\eta O| \phi \rangle.
\end{equation}
Indeed, since the metric operator $\eta$ and the generator $H$ satisfy the intertwining relation $\eta H = H^\dagger \eta$, we have
\begin{align}\nonumber
\langle U(t)\psi , U(t) \phi \rangle &\equiv  \langle \psi |U(t)^{\dagger}\eta U(t)| \phi \rangle=  \langle \psi |\eta U(t)^{-1}U(t)| \phi \rangle\\ &= \langle \psi |\eta| \phi \rangle \equiv \langle \psi , \phi \rangle .
\end{align}

We now briefly discuss the equivalence with the Isospectral Counterpart Hamiltonian $h$; see Ref.~\cite{Mostafazadeh2002b}. $\eta^{1/2}$ defines a similarity transformation for $\mathcal{PT}$-symmetric NHHs $H$ with real eigenvalues (in the $\mathcal{PT}$-symmetric phase) to Hermitian isospectral Hamiltonians $h$, by $h = \eta^{1/2}H\eta^{-1/2}$. 
By endowing the Hilbert space with the pseudo-metric operator $\eta$ (which is not unique), the $PT$-symmetric observable can be re-expressed in terms of the Hermitian Hamiltonian:
\begin{align}
\langle O(t) \rangle^{\eta}_{|\psi\rangle} 
&=\langle \psi| \eta e^{i H t}O e^{-i H t}|\psi \rangle /  \langle \psi|  \eta |\psi \rangle \nonumber\\
&=\langle \psi| \eta \eta^{-1/2} e^{i h t}\eta^{1/2}O \eta^{-1/2}e^{-i h t}\eta^{1/2}|\psi \rangle \nonumber\\ 
&\quad \quad \times 1/  \langle \psi|  \eta^{1/2}\eta^{1/2} |\psi \rangle \nonumber\\
&=\langle \phi|  e^{i h t}\widetilde{O} e^{-i h t}|\phi \rangle /  \langle \phi|\phi \rangle \nonumber\\
&= \langle \phi | \widetilde{O}(t)|\phi \rangle.
\end{align}
Where  $\widetilde{O}\equiv \eta^{1/2}O \eta^{-1/2}$ is Hermitian provided $O$ is $\eta$-pseudo Hermitian \citep{Meden2023}, and $|\phi\rangle \equiv \eta^{1/2}|\psi\rangle$. 

\bibliography{refs} 

\end{document}